\documentclass{aa} 
\usepackage[switch]{lineno} 
\usepackage{lipsum} %Creates example text
\usepackage[version=4]{mhchem}
\usepackage{graphicx}
\usepackage[colorlinks=true,citecolor=blue,linkcolor=blue,urlcolor=blue]{hyperref}
\usepackage{txfonts}
\usepackage{soul}
\usepackage{subcaption}
\usepackage{makecell}
\begin{document}
%\linenumbers
   %\title{Mini Photochemical Networks with Net Reactions (Suitable for GCMs)}
   \title{A Mini-Chemical Scheme with Net Reactions for 3D GCMs I.: Thermochemical Kinetics}
   \author{Shang-Min Tsai\inst{1},  Elspeth K. H. Lee\inst{2}, Raymond Pierrehumbert\inst{1}} %Franck Selsis(?),
    \institute{Atmospheric, Ocean, and Planetary Physics, Department of Physics, University of Oxford, UK\\
               \email{shang-min.tsai@physics.ox.ac.uk}
         \and
             Center for Space and Habitability, University of Bern, Gesellschaftsstrasse 6, CH-3012 Bern, Switzerland.}
\date{}

%{}{}{}{} 

% 5 {} token are mandatory
 
  \abstract
  % context heading (optional)
   {Growing evidence has indicated that the global composition distribution plays an indisputable role in interpreting observational data. 3D general circulation models (GCMs) with a reliable treatment of chemistry and clouds are particularly crucial in preparing for the upcoming observations. In the effort of achieving 3D chemistry-climate modeling, the challenge mainly lies in the expensive computing power required for treating a large number of chemical species and reactions.}
  % methods heading (mandatory)
  {Motivated by the need for a robust and computationally efficient chemical scheme, we devise a mini-chemical network with a minimal number of species and reactions for \ce{H2}-dominated atmospheres.}
  % Method
 {We apply a novel technique to simplify the chemical network from a full kinetics model -- VULCAN by replacing a large number of intermediate reactions with net reactions. The number of chemical species is cut down from 67 to 12, with the major species of thermal and observational importance retained, including \ce{H2O}, \ce{CH4}, CO, \ce{CO2}, \ce{C2H2}, \ce{NH3}, and HCN. The size of the total reactions is greatly reduced from $\sim$ 800 to 20. The mini-chemical scheme is validated by verifying the temporal evolution and benchmarking the predicted compositions in four exoplanet atmospheres (GJ 1214b, GJ 436b, HD 189733b, HD 209458b) against the full kinetics of VULCAN.
}%The mini-network encompasses major species of thermal and observational importance, such as \ce{H2O}, \ce{CH4}, CO, \ce{CO2}, \ce{C2H2}, \ce{NH3}, HCN etc., but only consists of 10 forward reactions.
  % results heading (mandatory) % across 500 K $\leq T \leq$ 3000 K and 10$^{-6}$ bar $\leq P \leq$ 10$^3$ bar 
{The mini-network reproduces the chemical timescales and composition distributions of the full kinetics well within an order of magnitude for the major species in the pressure range of 1 bar -- 0.1 mbar across various metallicities and carbon-to-oxygen (C/O) ratios.
}
% Conclusion
{We have developed and validated a mini-chemical scheme using net reactions to significantly simplify a large chemical network. The small scale of the mini-chemical scheme permits simple use and fast computation, which is optimal for implementation in a 3D GCM or a retrieval framework. We focus on the thermochemical kinetics of net reactions in this paper and address photochemistry in a follow-up paper.
}

       \titlerunning{A Mini-Chemical Scheme with Net Reactions for 3D GCMs} % short title
       \authorrunning{Tsai et al.}
   \maketitle

%-------------------------------------------------------------------

\section{Introduction} % There is urgent neeed to explore the 3D effects of disequilbirium chemistry in the era of XXX.
The field of exoplanet research is now entering the stage of probing the spatial distribution of atmospheric composition \citep{Venot2018,Ehrenreich2020}. The upcoming observatories, such as the James Webb Space Telescope (JWST; \cite{Venot2020,Drummond2020}) and the Atmospheric Remote-sensing Infrared Exoplanet Large-survey (Ariel; \citep{Moses2021,Tinetti2021}), will have the ability to provide accurate spectral data and map out the compositional variation across the globe of the planet. Chemical kinetics models \citep[e.g.,][]{Kasting1979,Yung1984,Moses11,Venot12,Hu2012,Miguel2014,Karan2019,Hobbs19,Tsai2021} have played an instrumental role in understanding the fundamental processes that shape the atmospheric compositions. However, these models are commonly limited to a 1D column approach. Studies using 3D models, such as \cite{Drummond2018,Mendonca2018a,Drummond2020}, have demonstrated the importance of horizontal transport on tidally-locked exoplanets. Considering the effects of global circulation
is critical in understanding the chemical and thermal feedbacks and interpreting phase-resolved observational data. In addition, retrieval works \citep{Taylor2020,Feng2020,Irwin2020,Pluriel2020,Pluriel2022} have shown that atmospheric retrievals can suffer biases when neglecting the 3D nature of the planets. Pseudo-2D models employing a rotating 1D-column have started to emerge \citep{Agundez2014,Venot2020b,Baeyens2021,Moses2021,Roth2021} and significantly improved the lack of horizontal interconnection in 1D models, but the circulation is considerably simplified with a globally uniform jet and the chemical-radiative feedback is excluded. A self-consistent 3D GCM coupling chemistry, radiation, and circulation is desired to study their interactions in depth and to be in position for the prospective observations.

The endeavor of coupling the gaseous chemistry to a 3D general circulation model (GCM) began with the chemical relaxation method \citep{Cooper2006} and later followed by \cite{Drummond2018,Mendonca2018a}. The relaxation method \citep{Cooper2006,tsai18} is analogous to Newtonian cooling as the radiative heating in idealized GCMs, which is the simplest to implement and least computationally demanding to implement in a 3D model. However, it is challenging to generalize the relaxation method to incorporate photochemistry since a predetermined photochemical equilibrium state is required in principle. Motivated by the need for a more efficient scheme, \cite{Venot2019} developed a reduced chemical scheme cutting down 105 species and about 2000 reactions from the original network \citep{Venot12} to 30 species and 362 reactions. The size of the scheme from \cite{Venot2019} is close to that in \cite{tsai17}, which is probably the size limit to maintain accuracy for conventional kinetics. \cite{Venot2020} further updated the methanol (\ce{CH3OH}) chemistry and added acetylene (\ce{C2H2}) to extend the valid domain for warm carbon-rich atmospheres. However, the C-H-O network in \cite{tsai17} does not include nitrogen chemistry, whereas the new reduced scheme in \cite{Venot2020} now involves 44 species and 582 reactions, and photochemistry is not considered in all of the above schemes. \cite{Chen2021} have recently applied an Earth-based chemistry-climate model (CCM) \citep{Marsh2013} employing the MOZART chemical module \citep{Kinnison2007} to explore the impact of stellar flares. The setup provides insights into potential Earth twins but is restricted to atmospheres with Earth-like compositions. Compared to the development of atmospheric chemistry modules for Earth-climate models \citep[e.g.,][]{Kinnison2007,Derwent2021} and the progress of simulating aerosols \citep{Lee2016,Lines2018,Steinrueck2021}, a robust chemical scheme with photochemistry capacity is still lacking and timely needed in exoplanet science. 

In this work, we present a novel design of the chemical scheme aiming to tackle the aforementioned problems. The chemical network is composed of a few elementary reactions that treat radical species and a handful of net reactions that greatly reduce the kinetics mechanisms. Our C-H-N-O thermochemical network without photochemistry consists of only 12 species and 10 forward reactions. The scheme is validated for a wide range of temperature, pressure, elemental abundances, and has the capacity to include photochemistry. The scheme is suitable for applications that require minimal computing time such as 3D GCMs and atmospheric retrievals. We will focus on the method and validation of the net-reaction mechanisms in this paper and address photochemistry in a follow-up paper. 

\section{Method}
\begin{figure}
   \centering
   \includegraphics[width=\hsize]{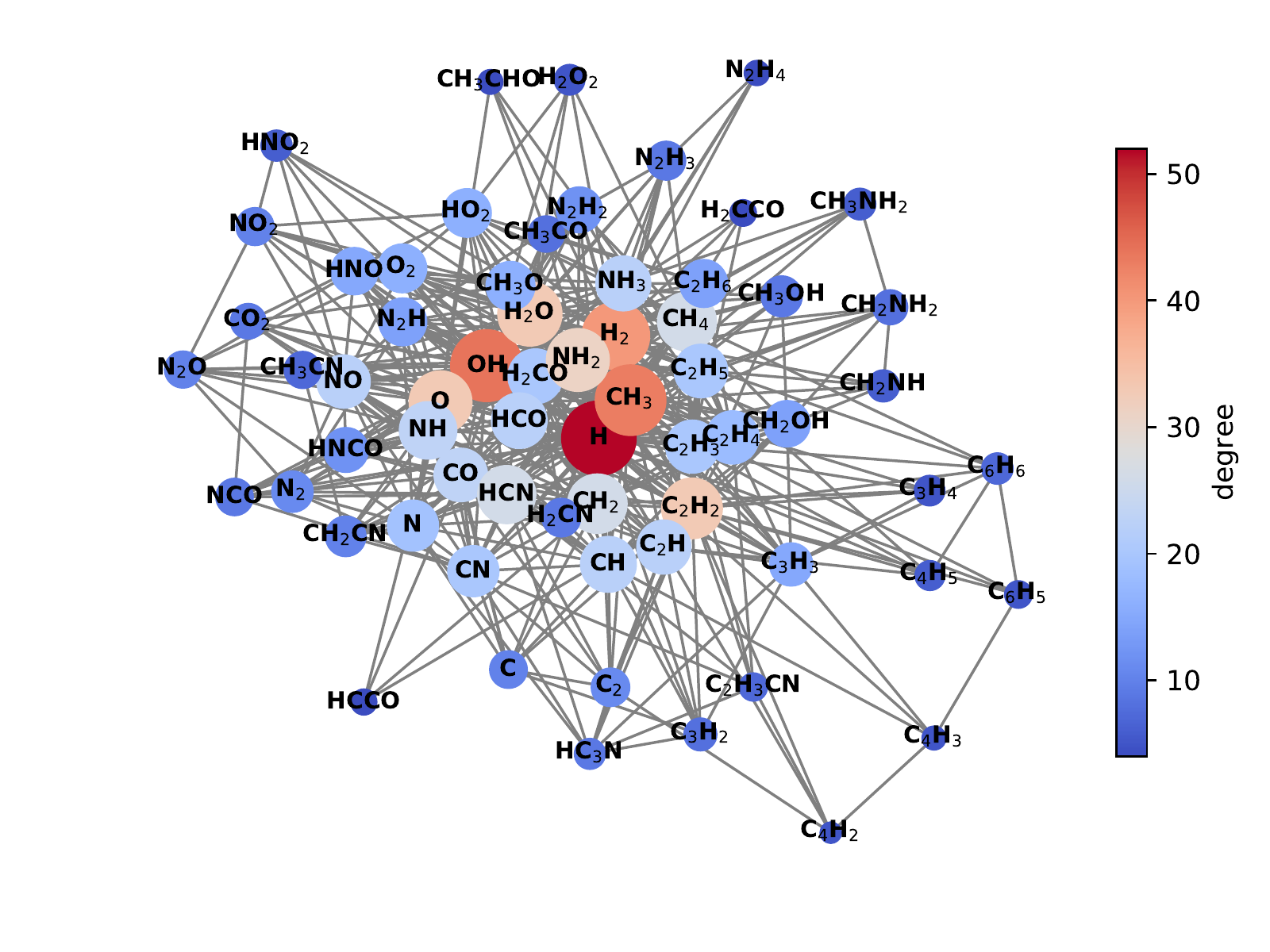}
   \includegraphics[width=0.95\hsize]{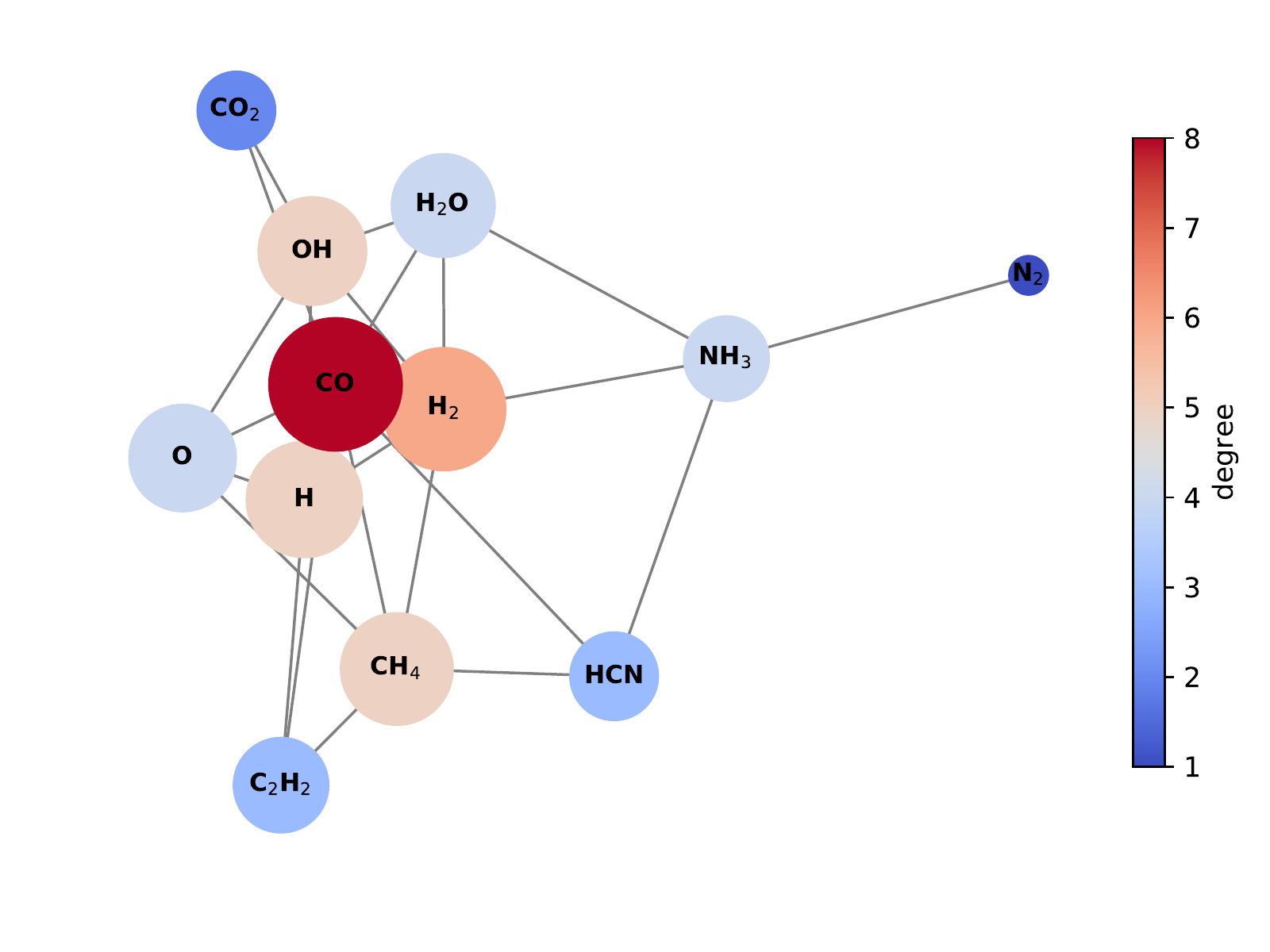}
    \caption{Visualization of the full chemical network from VULCAN (top) and the mini-network (bottom). Each species is represented by a node with the color varies with the degree and the size varies with centrality. The shorter length of the edges (lines) indicates faster rates between two species (not to linear scale). The graphs are for T = 1000 K, P = 1 bar and chemical equilibrium composition.}\label{graph}
   \end{figure}
\subsection{Making use of net reactions}
The principal mechanisms governing chemical species are often understood by the associated cycles (or referred as schemes, e.g., \cite{Moses11} and pathways, e.g., \cite{Venot2020,Tsai2021}). The chemical conversions generally consist of more than one intermediate reaction step, e.g., the ozone cycle on Earth \citep[e.g.,][]{Jacob2011} and the \ce{CH4}--CO interconversion on Jupiter \citep{Prinn1977,Visscher2010} and brown dwarfs \citep{Zahnle2014}. It is essential in kinetics simulations to include all reactions relevant to the application, supplied with correct rate coefficients. The chemical cycles would naturally emerge as an outcome of this bottom-up approach. In this work, we followed \cite{tsai18} and applied Dijkstra's algorithm \citep{Dijkstra} to identify the fastest conversion pathways for different atmospheric conditions systematically. Taking one of the \ce{CH4}--CO conversion pathways in a warm \ce{H2} atmosphere as an example:
\begin{eqnarray}
\begin{aligned}
\ce{ CH4 + H &-> CH3 + H2}\\
\ce{ CH3 + OH &->[M] CH3OH}\\
\ce{ CH3OH + H &-> CH3O + H2}\\
\ce{ CH3O &->[M] H2CO + H}\\
\ce{ H2CO + H &-> HCO + H2}\\
\ce{ HCO &->[M] H + CO}\\
\ce{ H + H2O &-> OH + H2} \\
\ce{ H2 &->[M] 2H}\\
\noalign{\vglue 5pt} 
\hline
\noalign{\vglue 5pt} 
\mbox{net} : \ce{CH4 + H2O &-> CO + 3H2},\label{ch4-co-pathway}
\end{aligned}
\end{eqnarray}
where \ce{CH4 + H2O -> CO + 3H2} with unspecified rate coefficient is simply a mathematical ``summary'' of the above eight reactions that compose the pathway sequence. An attractive property of the pathway is that the overall timescale of the conversion is controlled by the slowest reaction, i.e. the rate-limiting step \citep{Moses11,tsai18}, which is readily determined once the pathway is identified. The rate-limiting step per se contains sufficient information for computing the rate of change without getting into the details of each elementary reaction.  We construct the network with the top-down design, as opposed to the bottom-up structure in conventional kinetics. The crux of the mini-network is to replace hundreds of elementary reactions in a full network with just a few net reactions. The effective rates of these net reactions are subsequently determined by the corresponding rate-limiting steps. % (The rates of net reactions are usually unknown while only those of the individual reactions are measured or computed. However, ) 

% For example, such as (\ref{ch4-co-pathway}) 
We emphasize that the pathways and their rate-limiting steps depend strongly on the atmospheric temperature, pressure, and elemental abundances. As a result, the rate coefficient of these net reactions can no longer be expressed by the modified Arrhenius equation, which is generally a function solely depends on temperature\footnote{except for some reactions that require a third-body collision and hence have pressure dependence}, i.e. $k = A T^{b} \exp{\left( - \frac{E}{T} \right)}$. Instead, the effective rate coefficients of a schematic net reaction \ce{A + B -> C + D} dictated by the rate-limiting step is expressed as
\begin{equation}
k = \frac{\textrm{rate}_{\textrm{RLS}}}{[A][B]}\label{net-rate} 
\end{equation}
where $\textrm{rate}_{\textrm{RLS}}$ is the reaction rate (molecules cm$^{-3}$ s$^{-1}$) of the rate-limited step in the entailed pathway, and [A], [B] the mole fraction of the reactants A, B. All quantities in (\ref{net-rate}) are evaluated in chemical equilibrium for the given temperature, pressure, and elemental abundances (see the discussion regarding adopting equilibrium abundances in \cite{tsai18}). Hence the rate coefficient (\ref{net-rate}) is now a function of temperature, pressure, and elemental abundances. We then derive the rate coefficients of the backward net reactions by reversing those of the forward reactions to ensure thermochemical equilibrium can be consistently achieved \citep{tsai17}, the same way as with elementary reactions.

Following the same example, (\ref{ch4-co-pathway}) presents the pathway of \ce{CH4}--CO conversion at $T$ = 1000 K and $P$ = 1 bar where \ce{ CH3 + OH ->[M] CH3OH} is the rate-liming step. At the same pressure but with the temperature increased to 1500 K, the pathway switches to
\begin{eqnarray}
\begin{aligned}
\ce{ CH4 + H &-> CH3 + H2}\\
\ce{ CH3 + OH &-> CH2OH + H}\\
\ce{ CH2OH &->[M] H2CO + H}\\
\ce{ H2CO + H &-> HCO + H2}\\
\ce{ HCO &->[M] H + CO}\\
\ce{ H + H2O &-> OH + H2} \\
\noalign{\vglue 5pt} 
\hline
\noalign{\vglue 5pt} 
\mbox{net} : \ce{CH4 + H2O &-> CO + 3H2},\label{ch4-co-pathway2}
\end{aligned}
\end{eqnarray}
where \ce{ CH3 + OH -> CH2OH + H} is now the rate-liming step. Accordingly, the rate coefficient of the net reaction \ce{CH4 + H2O -> CO + 3H2} at $T$ = 1000 K and $P$ = 1 bar is
\begin{equation}\label{net-rate1}
k = \frac{k_1[\ce{CH3}][\ce{OH}]\ce{M}}{[\ce{CH4}][\ce{H2O}]}, 
\end{equation}
while that at $T$ = 1500 K and $P$ = 1 bar is
\begin{equation}\label{net-rate2}
k = \frac{k_2[\ce{CH3}][\ce{OH}]}{[\ce{CH4}][\ce{H2O}]},
\end{equation}
where $k_1$ and $k_2$ are the rate coefficients of \ce{ CH3 + OH ->[M] CH3OH} and \ce{ CH3 + OH -> CH2OH + H}, respectively.

Based on the major pathways among the key molecules \citep{Moses11,tsai18,Venot2020,Tsai2021b}, we employ six essential net reactions to govern the main species in the C-H-N-O thermochemical kinetics. Firstly, \ce{CH4 + H2O -> CO + 3H2} and \ce{2NH3 -> N2 + 3H2} describe the \ce{CH4}--CO and \ce{NH3}--\ce{N2} interconversions, respectively. \ce{2CH4 -> C2H2 + 3H2} is identified as the main channel for \ce{C2H2} production at low temperature/high pressure and \ce{CO + CH4 -> C2H2 + H2 + O} at high temperature/low pressure, where CO is the main carbon-bearing molecule. Similarly, \ce{CH4 + NH3 -> HCN + 3H2} and \ce{CO + NH3 -> HCN + H2O} are employed for HCN production at low temperature/high pressure and high temperature/low pressure, respectively. 
Additionally, four elementary reactions involve fast-reacting radicals: OH, H, and O are included to complete the mini-network. Specifically, \ce{OH + H2 -> H2O + H} is a key reaction for the formation of water in a hydrogen-rich environment \citep[e.g.,][]{Liang2003,Tsai2021b}. \ce{OH + CO -> H + CO2} is responsible for the interconversion between CO and \ce{CO2} \citep[e.g.,][]{Yung1999,Gao2015}. The above two reactions are necessary to correctly compute \ce{H2O} and \ce{CO2}. Lastly, \ce{O + H2 -> OH + H} contributes to tracking atomic O and \ce{H + H ->[M] H2} to hydrogen dissociation and recombination, which are included to be in position for the implementation with photochemistry.

All the elementary and net reactions employed in our mini-network are listed in Table \ref{tab:network}, which encompass 12 species: H, \ce{H2}, OH, \ce{H2O}, CO, \ce{CO2}, O, \ce{CH4}, \ce{C2H2}, \ce{NH3}, \ce{N2}, HCN. The topology of the mini-network and the full network that the mini-network is condensed from is illustrated in Fig. \ref{graph}, where the degree means the number of reaction connections to other species and the eigenvector centrality measures the influence of the species by taking into account both quantity (number of reaction links) and quality (rates of reactions and connections to reactive species). In the mini-network, H lost its high centrality in the full network since most of the elementary reactions involving H are now concealed in the net reactions. Similarly, the fast cycles between \ce{CH3} and \ce{CH4} and those between \ce{NH3} and \ce{NH2} are implicitly packed in the net reactions. Our mini-network keeps most of the major species with the highest centrality in the full kinetics, except for excluding \ce{CH3} and \ce{NH2} for simplicity and including \ce{N2} as a major nitrogen-bearing molecule.

%({\bf come back to CH3 in the discussion?})
% {\bf  (see \cite{Moses11,Venot2020,Tsai2021b} for more on the corresponding pathways of the net reactions in different atmospheric conditions}

\begin{figure}
   \centering
   \includegraphics[width=\hsize]{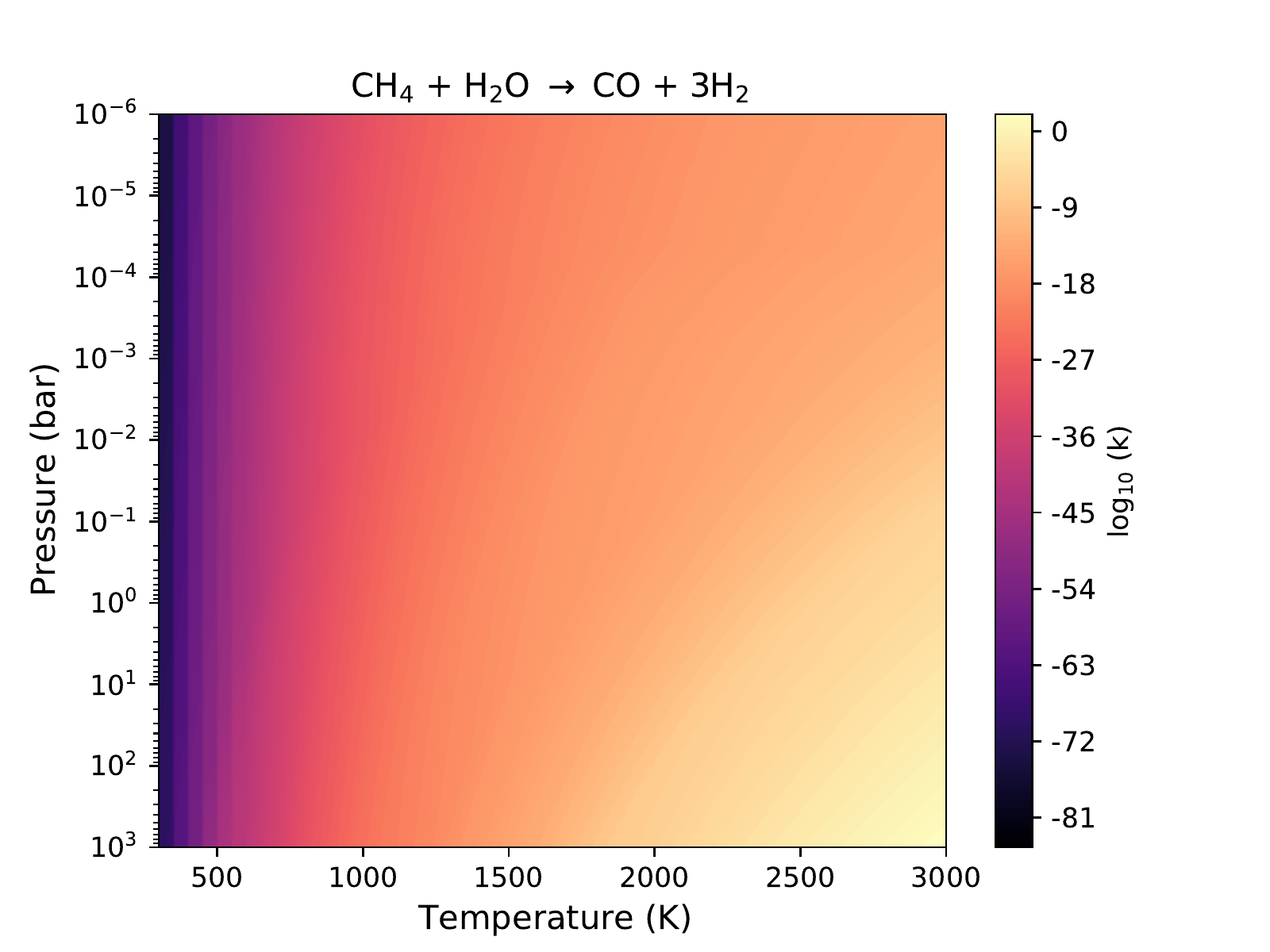}
    \caption{The effective rate constants (cm$^3$ molecules$^{-1}$ s$^{-1}$) for the net reaction \ce{CH4 + H2O -> CO + 3H2}.}\label{fig:rates} 
\end{figure}
   
%Lastly, \ce{CH4 + CH4 -> C2H6 + H2} and \ce{ C2H2 + 2H2 -> C2H6} complete the main hydrocarbon mechanism. 

\begin{table*}[h]
\begin{minipage}{\linewidth}
\begin{center}
\caption{List of elementary and net reactions included in the C-H-N-O chemical kinetics. The backward reactions are reversed numerically with thermodynamic data \citep{tsai17}}
%(backward reactions with even indexes are reversed numerically with thermodynamic data) and rate coefficients (cm$^3$ s$^{-1}$ for bimolecular reactions and s$^{-1}$ for k$_0$) of }
%\tiny
\begin{tabular}{l|l|l}
%\hline
Elementary reaction & Rate Coefficient (cm$^3$ molecules$^{-1}$ s$^{-1}$) & Reference\\
\hline
\ce{OH + H2 -> H2O + H} & 3.57 $ \times 10^{-16}$ $T^{1.52}$ $\exp(-1740/T)$ & 
\cite{R1}\\
\ce{OH + CO -> H + CO2} & 1.05 $ \times 10^{-17}$ $T^{1.5}$ $\exp(250/T)$ & \cite{Baulch1992}\\
\ce{O + H2 -> OH + H} & 8.52 $ \times 10^{-20}$ $T^{2.67}$ $\exp(-3160/T)$ & \cite{Baulch1992}\\
\ce{H + H ->[M] H2} & \makecell[l]{$k_0$\footnote{low-pressure limit (cm$^6$ molecules$^{-2}$ s$^{-1}$)} = 2.7 $\times 10^{-31}$ $T^{-0.6}$\\$k_{\infty}$\footnote{high-pressure limit (cm$^3$ molecules$^{-1}$ s$^{-1}$)} = $3.31 \times 10^{-6}$ $T^{-1}$} & \makecell[l]{\cite{RH2-low}\\\cite{RH2-high}}\\ %\makecell[l]
\hline
\multicolumn{2}{c}{Net reaction}\\ %  & Note
\hline
%\multicolumn{3}{|c|}{Net reactions}\\
\multicolumn{2}{c}{\ce{CH4 + H2O -> CO + 3H2}}\\      
\multicolumn{2}{c}{\ce{2CH4 -> C2H2 + 3H2}}\\
%\multicolumn{3}{l}{\ce{2CH4 + CH4 -> C2H6 + H2}}\\
%\multicolumn{3}{l}{\ce{C2H2 + 2H2 -> C2H6}}\\ 
\multicolumn{2}{c}{\ce{CO + CH4 -> C2H2 + H2O}}\\       
\multicolumn{2}{c}{\ce{2NH3 -> N2 + 3H2}}\\ 
\multicolumn{2}{c}{\ce{CH4 + NH3 -> HCN + 3H2}}\\    
\multicolumn{2}{c}{\ce{CO + NH3 -> HCN + H2O}}\\     
\hline
\end{tabular}\label{tab:network}
\end{center}
\end{minipage}
\end{table*}

% For the C-H-N-O chemical kinetics, we utilize the following 8 net reactions:
% \ce{CH4 + H2O -> CO + 3H2}  
% \ce{CH4 + CH4 -> C2H2 + 3H2}
% \ce{CH4 + CH4 -> C2H6 + H2} 
% \ce{CO + CH4 -> C2H2 + H2 + O}
% \ce{CH4 + NH3 -> HCN + 3H2}
% \ce{CO + NH3 -> HCN + H2O} 
% \ce{ C2H2 + 2H2 -> C2H6}.
% Additionally, 5 ``real-world" reactions governing OH or H radicals are included to complete the mini-network: % fundemental
% \ce{OH + H2 -> H2O + H }\\
% \ce{OH + CO -> H + CO2}\\
% \ce{O + H2 -> OH + H}\\
% \ce{OH + C -> CO + H}\\         
% (put in a table)\\
The numerical rate coefficients of the net reactions as a function of temperature and pressure for given elemental abundances can be generated using the full chemical scheme in advance and thus do not add extra computational cost when applying the mini-scheme.
Fig. \ref{fig:rates} illustrates the wide range of rate coefficients of the net reactions \ce{CH4 + H2O -> CO + 3H2} across temperatures and pressures for solar metallicity. We have tabulated the rate coefficients for temperatures and pressure in the range of 300--3000 K and 10${^3}$--10$^{-6}$ bar, for a grid of metallicities (0.1$\times$ solar, solar, 10$\times$ solar, 100$\times$ solar, 500$\times$ solar) and C/O ratios (C/O = 0.25, solar, C/O = 1, C/O = 2). The numerical tables of net reactions in the mini-chemical scheme are available in the supplementary files while a part of the table for \ce{CH4 + H2O -> CO + 3H2} is shown in Table \ref{tab:ex-rates} for demonstration. The application should be restricted to the tested range of elemental ratios, and since the conversion pathways that control the effective rate coefficients of the net reaction can be sensitive to the atmospheric condition, these rate coefficients should ideally be made from the first principle for a specific elemental abundance ratio\footnote{Please contact the author for a specific elemental abundance ratio not provided.}. We end this section by noting that although the same pathway analysis in \cite{tsai18} is applied, the crucial difference is that the chemical sources and sinks are approximated by a linear expansion in the chemical relaxation method in \cite{tsai18}, whereas exactly the same format of rate equations that allows nonlinear dynamics as the standard kinetics is utilized in this work.

\subsection{Validation Setup}
\subsubsection{0D evolution in time}
We set up a 0D kinetics model to compare the temporal evolution computed by the mini-network and that from the full C-H-N-O kinetics of VULCAN \citep{Tsai2021}\footnote{\url{https://github.com/exoclime/VULCAN/blob/master/thermo/NCHO_thermo_network.txt}}. The 0D model is initialized with prescribed gas mixtures at a fixed temperature and pressure, which evolves with time toward thermochemical equilibrium, analogous to the experimental setup of a cell for monitoring the evolution of the gas mixture \citep{Peng2014,Fleury2019}. The initial gas mixtures are \ce{H2}, He, \ce{CH4}, \ce{H2O}, \ce{NH3}, partitioned by solar elemental abundances, except that \ce{CH4} and \ce{NH3} are replaced by CO and \ce{N2}, respectively, in the \ce{CH4} and \ce{NH3} dominated regime (low temperatures and high pressures) to clearly show the changes in time. The C-, O-, N- bearing molecules are scaled accordingly when the metallicity varies, and we keep oxygen fixed when changing the C/O ratio.

% We take [\ce{CH4}]/[\ce{H2}] = 5$\times$10$^{-4}$, [\ce{H2O}]/[\ce{H2}] = 1$\times$10$^{-3}$, [\ce{NH3}]/[\ce{H2}] = 1$\times$10$^{-4}$, and [He]/[\ce{H2}] = 0.2 for solar-like elemental abudances and scale accordingly when the metllicity varies. 

%The initial mixtures for temperatures lower than 1000 K are ... and for T $>$ 1000 K are ... , both corresbonding to solar-like elemental abudances. 

% for temperatures lower than 1000 K are ... and for T $>$ 1000 K are
\begin{figure}[!h]
  \centering
  \includegraphics[width=\hsize]{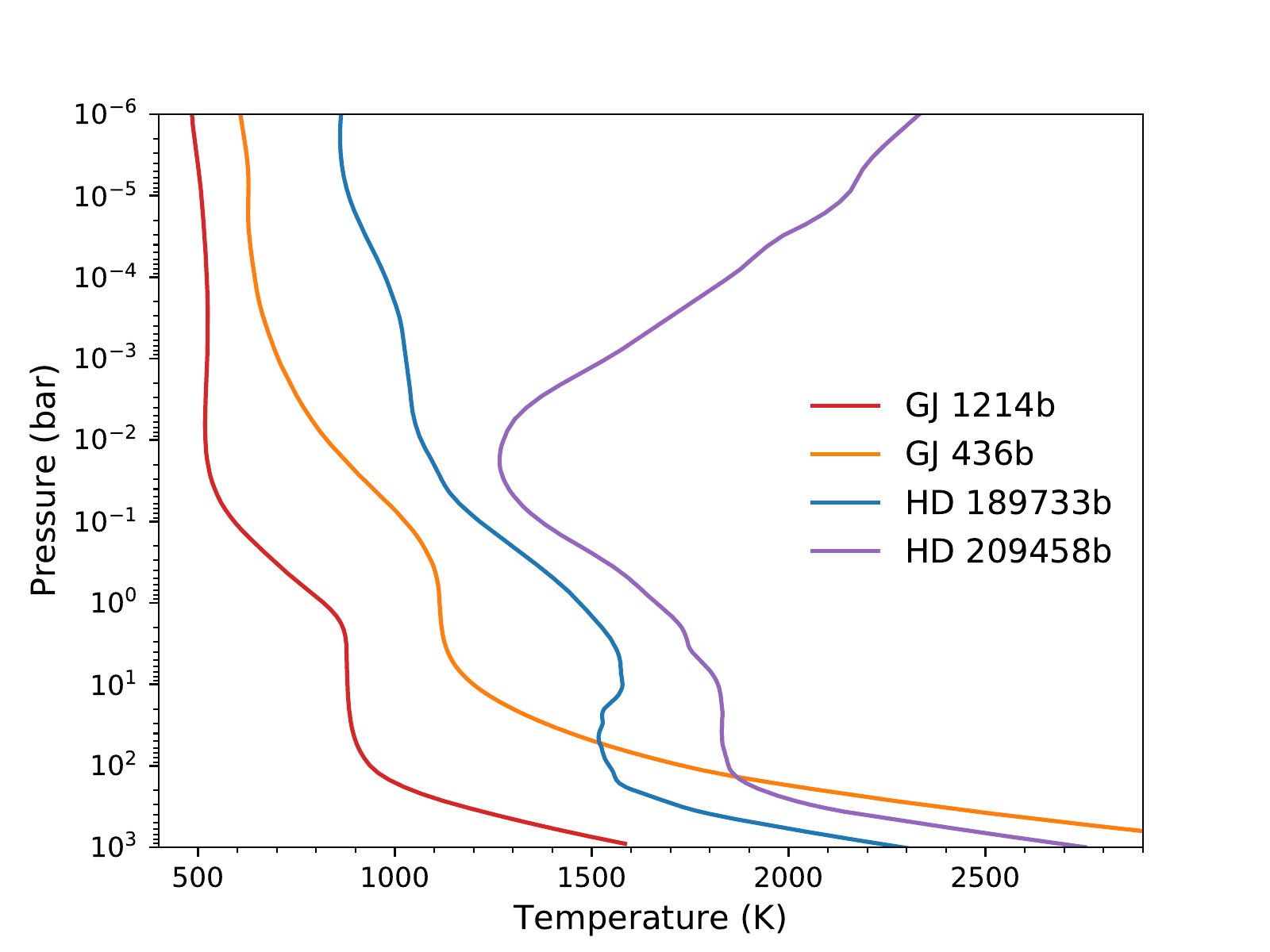}
    \caption{The adopted pressure-temperature profiles of GJ 1214b, GJ 436b, HD 189733b, and HD 209458b, for validating the mini-chemical scheme.}\label{TPs}
  \end{figure}
  
\subsubsection{1D vertical profiles}
Since the observable abundances in planetary atmospheres are usually governed by the transport-induced quenching process \citep[e.g.,][]{Baxter2021,Yui2021}, determining the quench levels \citep{vm11,Moses2014,tsai17} is the key aspect in 1D chemical kinetics modeling. To verify that our mini-chemical scheme can correctly reproduce the quenching behavior predicted in 1D models, we adopt the pressure--temperature ($P$--$T$) profiles of the planets GJ 1214b, GJ 436b, HD 189733b, and HD 209458b as inputs to validate the mini-network. These chosen atmospheres have equilibrium temperatures from about 500 K to 1700 K, representative of the vertical-mixing dominated regime. We compute the radiative-convective equilibrium temperature profiles of GJ 1214b and GJ 436b using the radiative-transfer model HELIOS \citep{Malik2019}, while those of HD 189733b and HD 209458b are taken from \cite{Moses11}. The inverted temperature profile of HD 209458b is adopted for validating the scheme with a thermal inversion and comparison with previous works \citep{tsai17,Venot2019}, but we note that emission observations show no evidence of thermal inversion of HD 209458b \citep[e.g.,][]{Diamond2014,Schwarz2015,Line2016}. All the $P$--$T$ profiles are shown in Fig. \ref{TPs}. We vary the uniform eddy diffusion coefficients ($K_{\textrm{zz}}$) from 10$^5$ to 10$^{11}$ cm$^2$/s, evenly spaced on a log10 scale, to explore diverse quench levels for each planet.
% Their equilibrium temperatures span from about 700 K to 1700 K and the temperature profiles are shown in Figure \ref{TPs}. 

% Use (the hot, warm, cool TPs in Tsai 2018 plus) HD189, GJ436b and 51 Eri b with ranges of Kzz to validate the thermochemistry. 

%Then HD189 and GJ436 for photochemistry.

%- explain that I also include multiple paths for the RLS 
%- first validate transport-induced quenching
%- then validate photochemistry
%- non-solar C/O and metallicity\\

% I find CH3 has a high degree. Can I justify eliminating it? Or, just show the tree diagram before and after simplification)

\begin{figure*}
   \centering
   \includegraphics[width=\hsize]{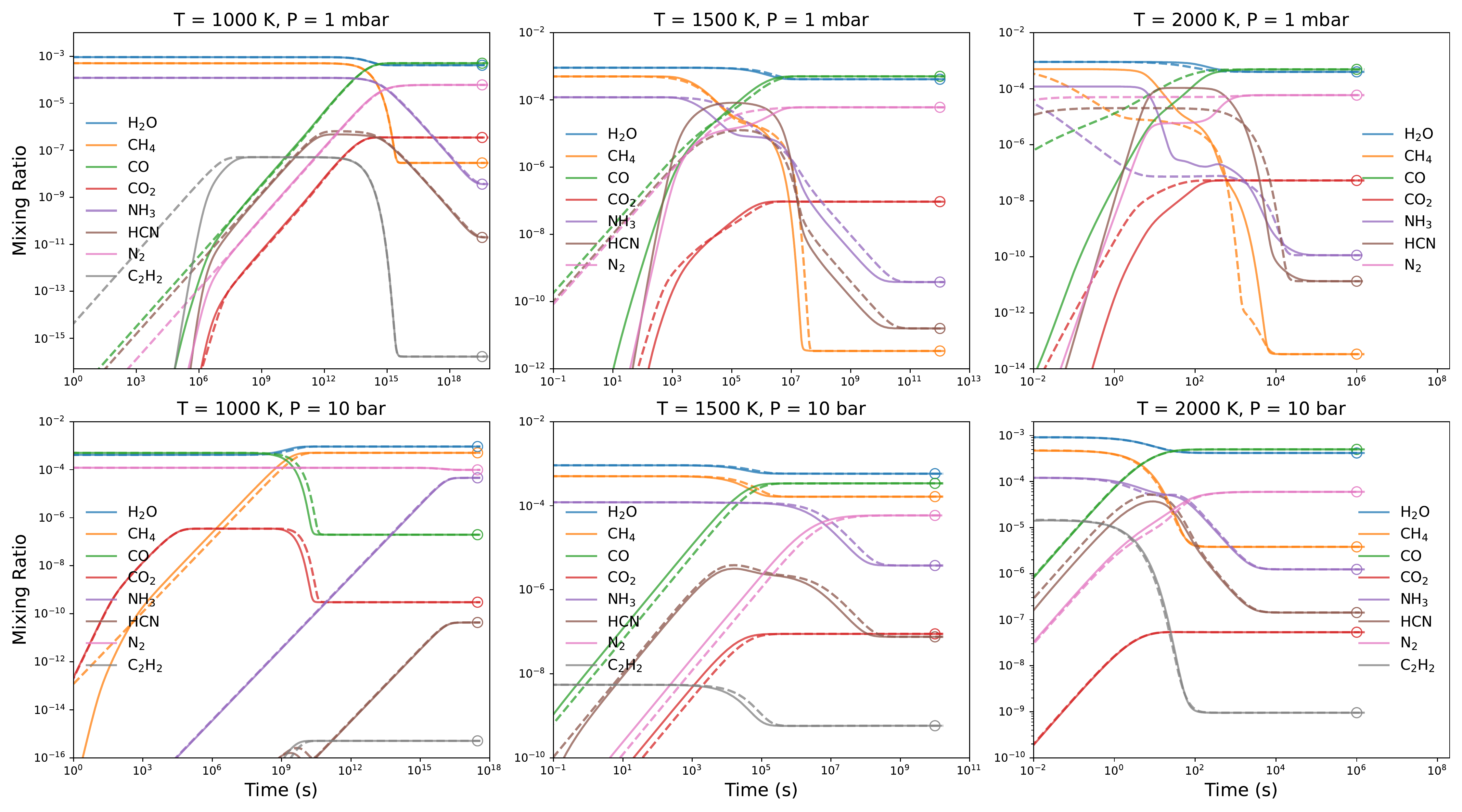}
    \caption{Time evolution of the main species in the 0D model computed with the full C-H-N-O kinetics (solid) and the mini-network (dashed) for solar metallicity with various temperatures and pressures. Open circles plotted in the end of each run indicate the thermochemical equilibrium abundances.}\label{0D-evo}
   \end{figure*}

% \begin{figure*}
% \centering     %%% not \center
% \subfig{\label{fig:a}\includegraphics[width=\columnwidth]{figs/CH4-all.pdf}\subcaption{\ce{CH4}}
% \subfig{\label{fig:b}\includegraphics[width=\columnwidth]{figs/CH4-all.pdf}}
% \caption{my caption}
% \end{figure*}

\begin{figure*}
  \centering
  \begin{subfigure}[t]{0.475\linewidth}
  \includegraphics[width=\columnwidth]{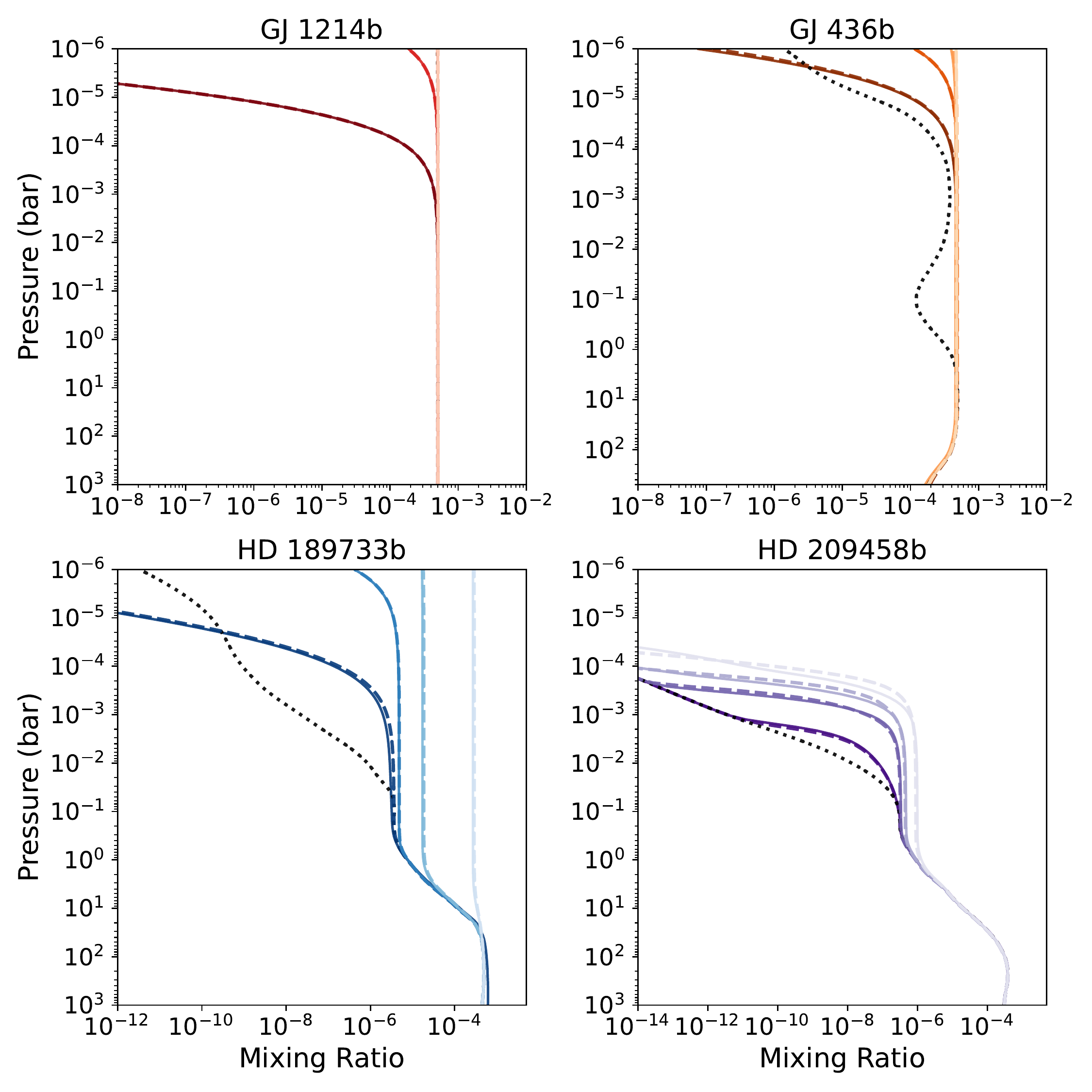}
  \subcaption{\ce{CH4}}
  \end{subfigure}
  \begin{subfigure}[t]{0.475\linewidth}
  \includegraphics[width=\columnwidth]{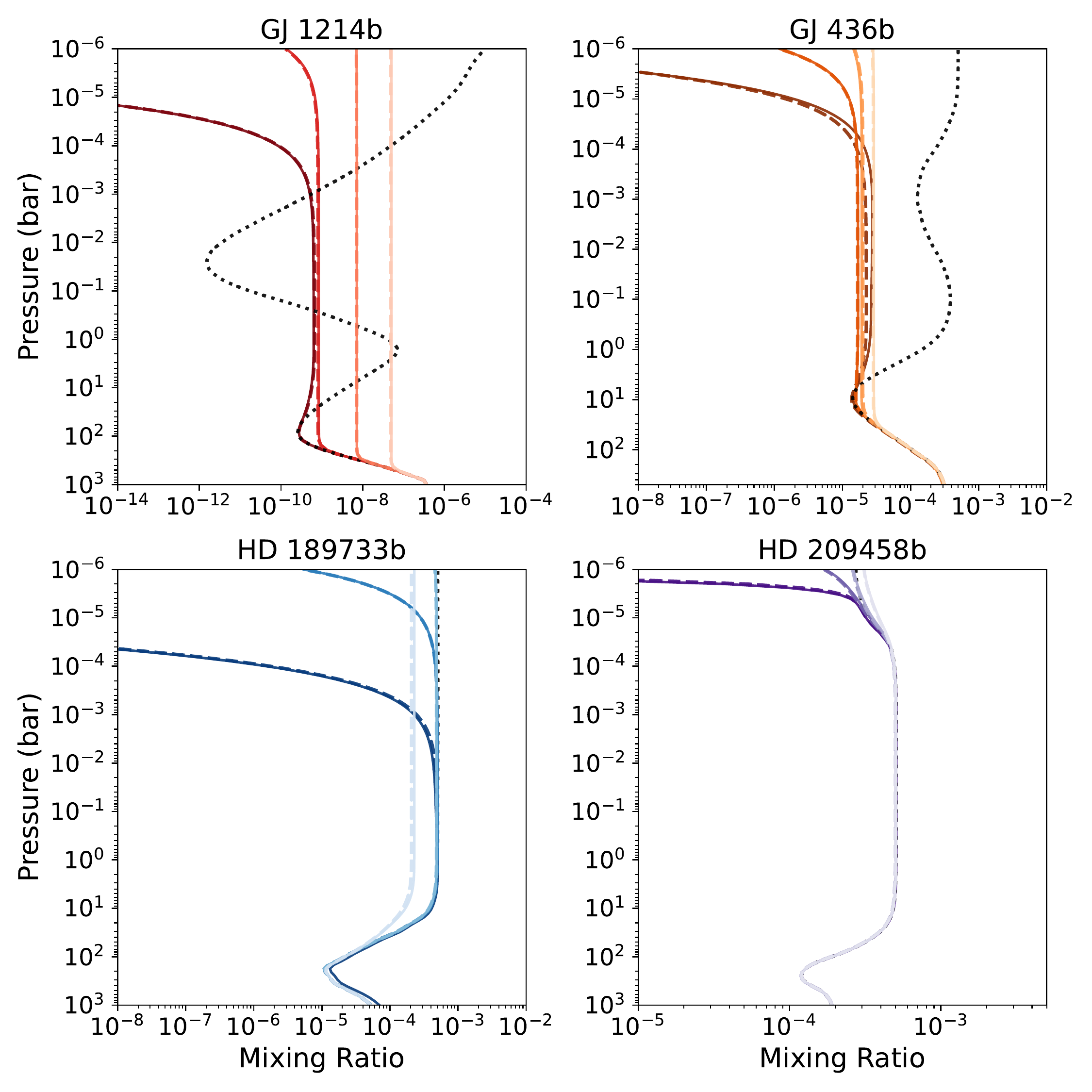}
  \subcaption{\ce{CO}}
  \end{subfigure}
  \begin{subfigure}[t]{0.475\linewidth}
  \includegraphics[width=\columnwidth]{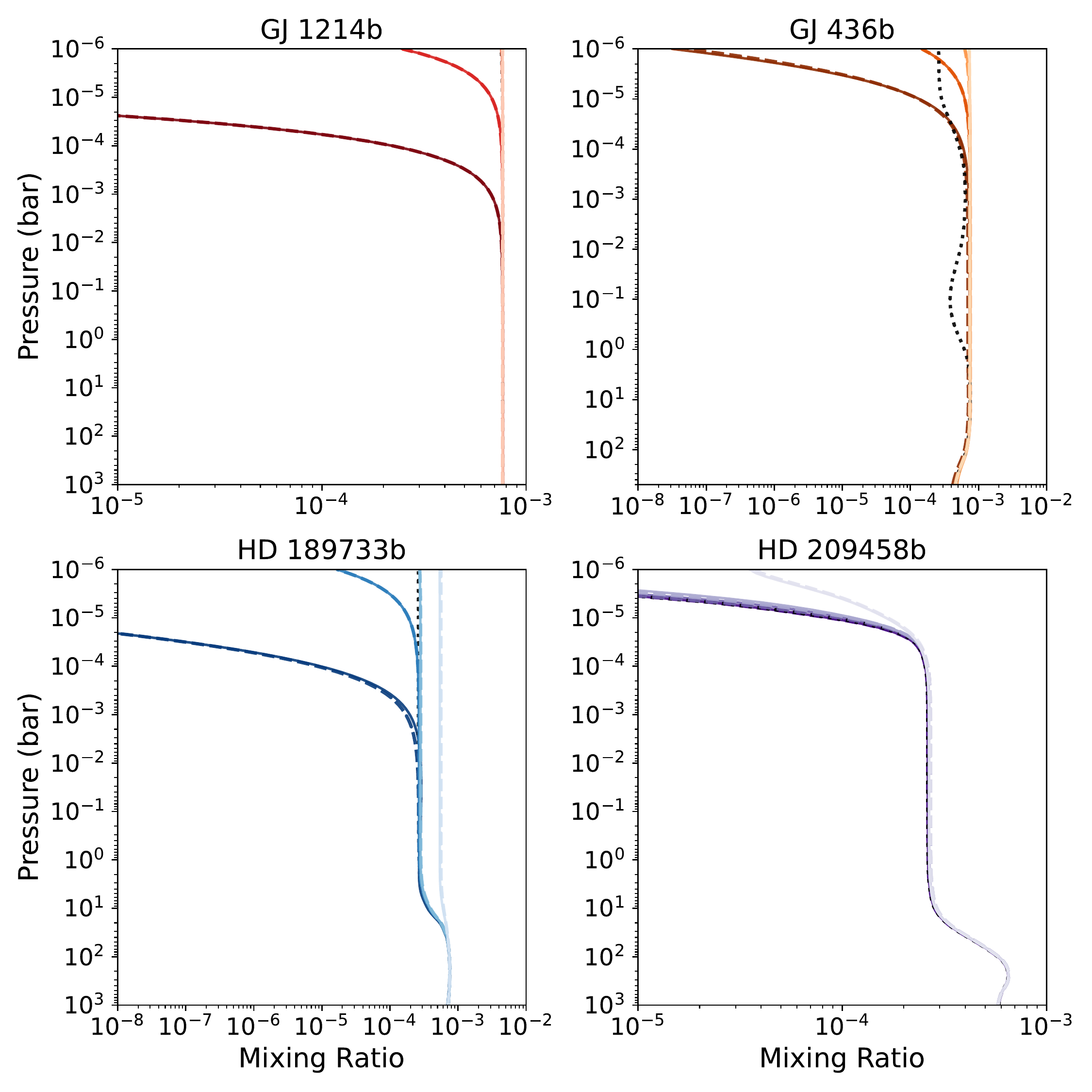}
  \subcaption{\ce{H2O}}
  \end{subfigure}
    \begin{subfigure}[t]{0.475\linewidth}
  \includegraphics[width=\columnwidth]{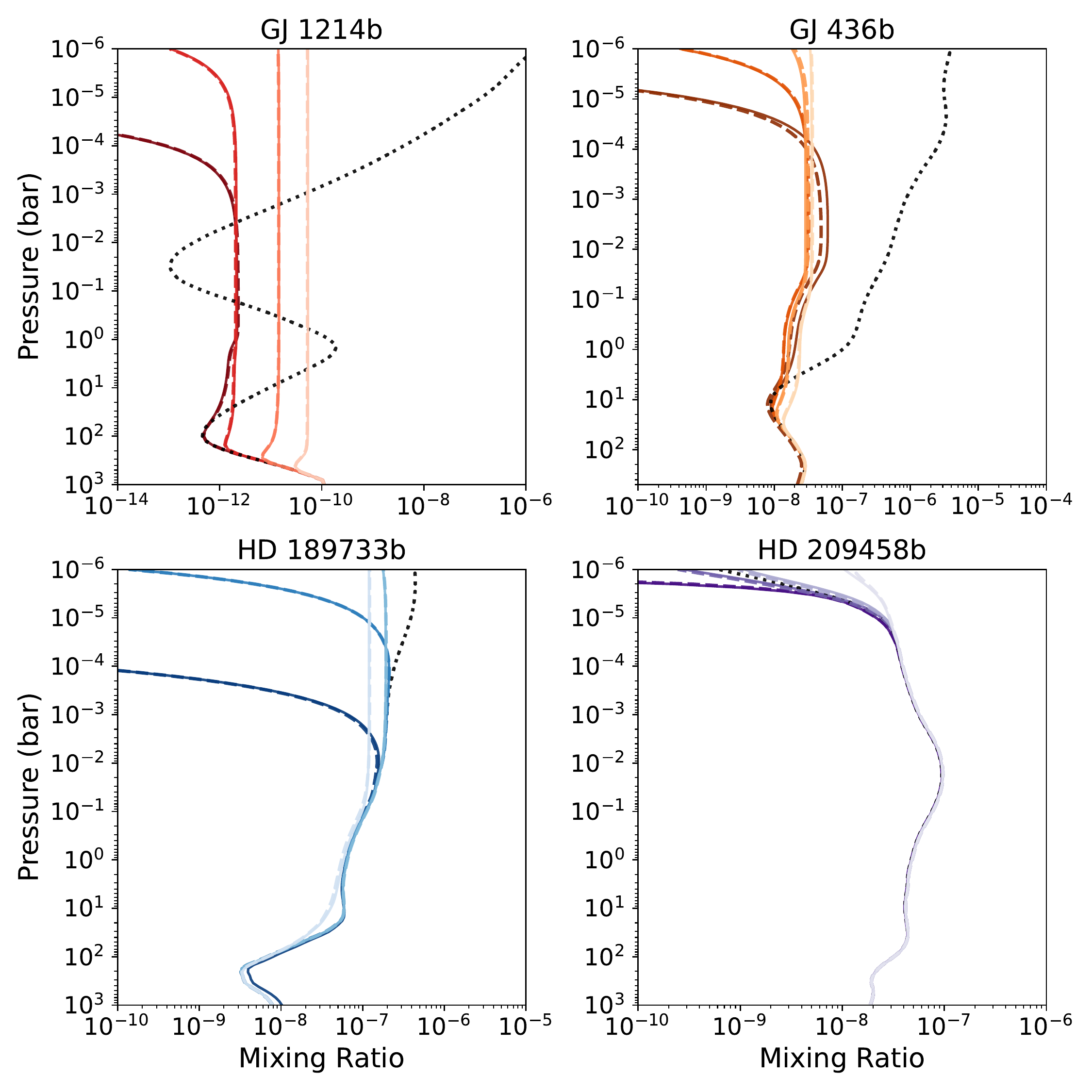}
  \subcaption{\ce{CO2}}
  \end{subfigure}
  \caption{The vertical composition distributions computed by the mini-chemical scheme (dashed) compared to those by the full chemical kinetics VULCAN (solid) for the four planets with temperature profiles in Fig. \ref{TPs}. The dark to light colors represent increasing vertical mixing (see the text for the specific values of $K_{\textrm{zz}}$ used).}\label{fig:1D} % Not showing the compositions with negligible mixing ratios ( $<$ 10$^{-12}$ between 1 and 10$^{-5}$ bar)
  \end{figure*}

\begin{figure*}
\ContinuedFloat
  \centering
%   \begin{subfigure}[t]{0.475\linewidth}
%   \includegraphics[width=\columnwidth]{figs/H-all-solar.pdf}
%   \subcaption{\ce{H}}
%   \end{subfigure}
    \begin{subfigure}[t]{0.475\linewidth}
  \includegraphics[width=\columnwidth]{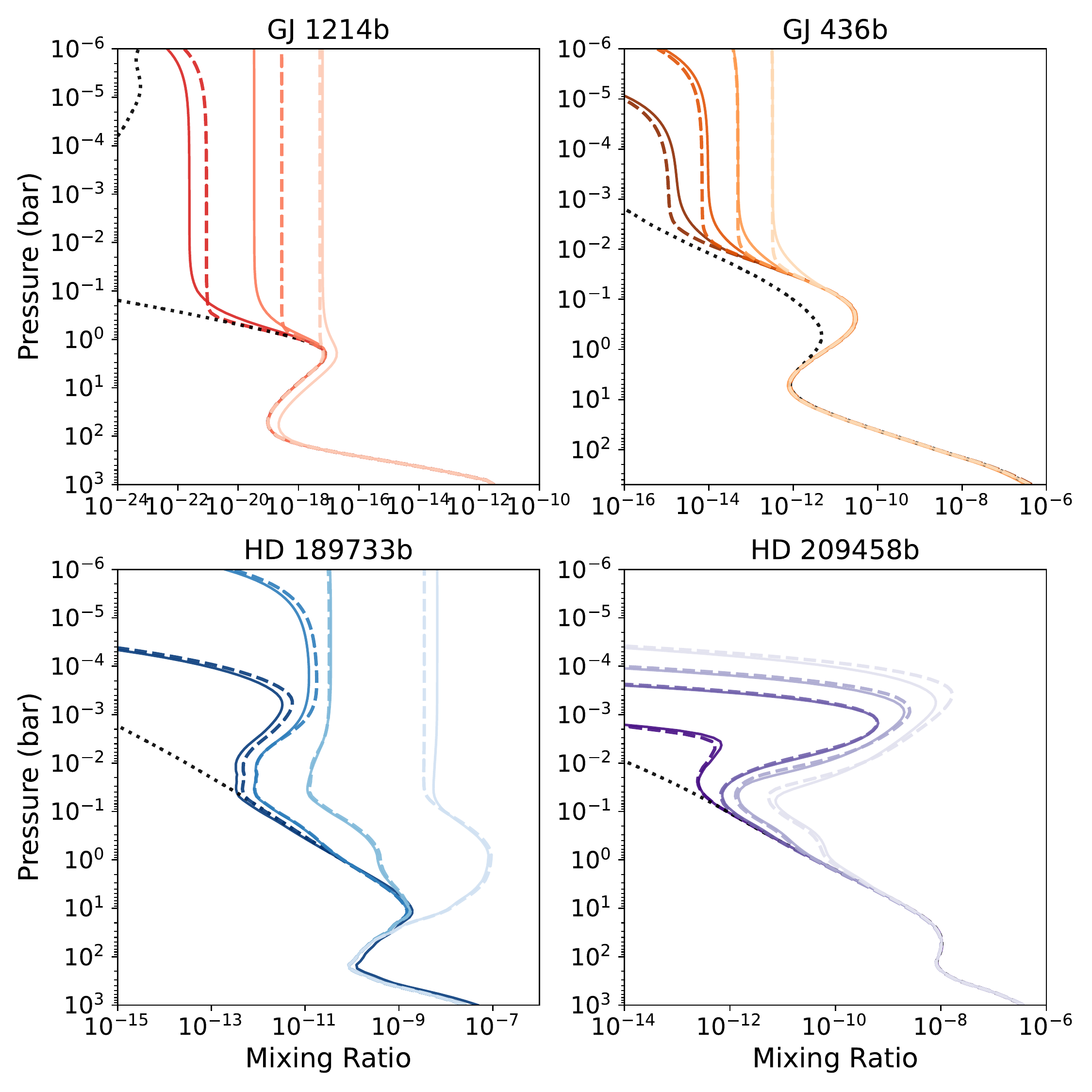}
  \subcaption{\ce{C2H2}}
  \end{subfigure}
  \begin{subfigure}[t]{0.475\linewidth}
  \includegraphics[width=\columnwidth]{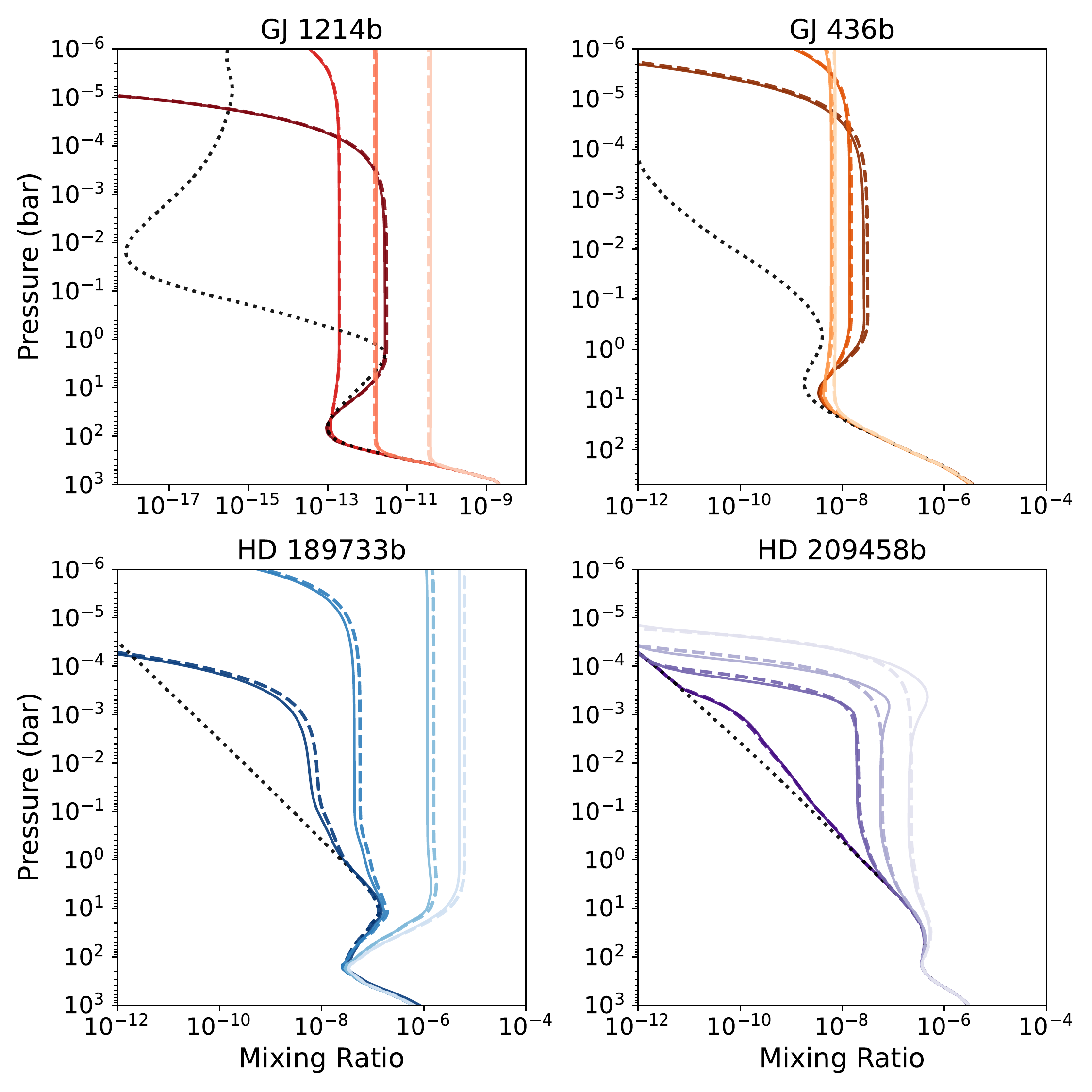}
  \subcaption{\ce{HCN}}
  \end{subfigure}
  \begin{subfigure}[t]{0.475\linewidth}
  \includegraphics[width=\columnwidth]{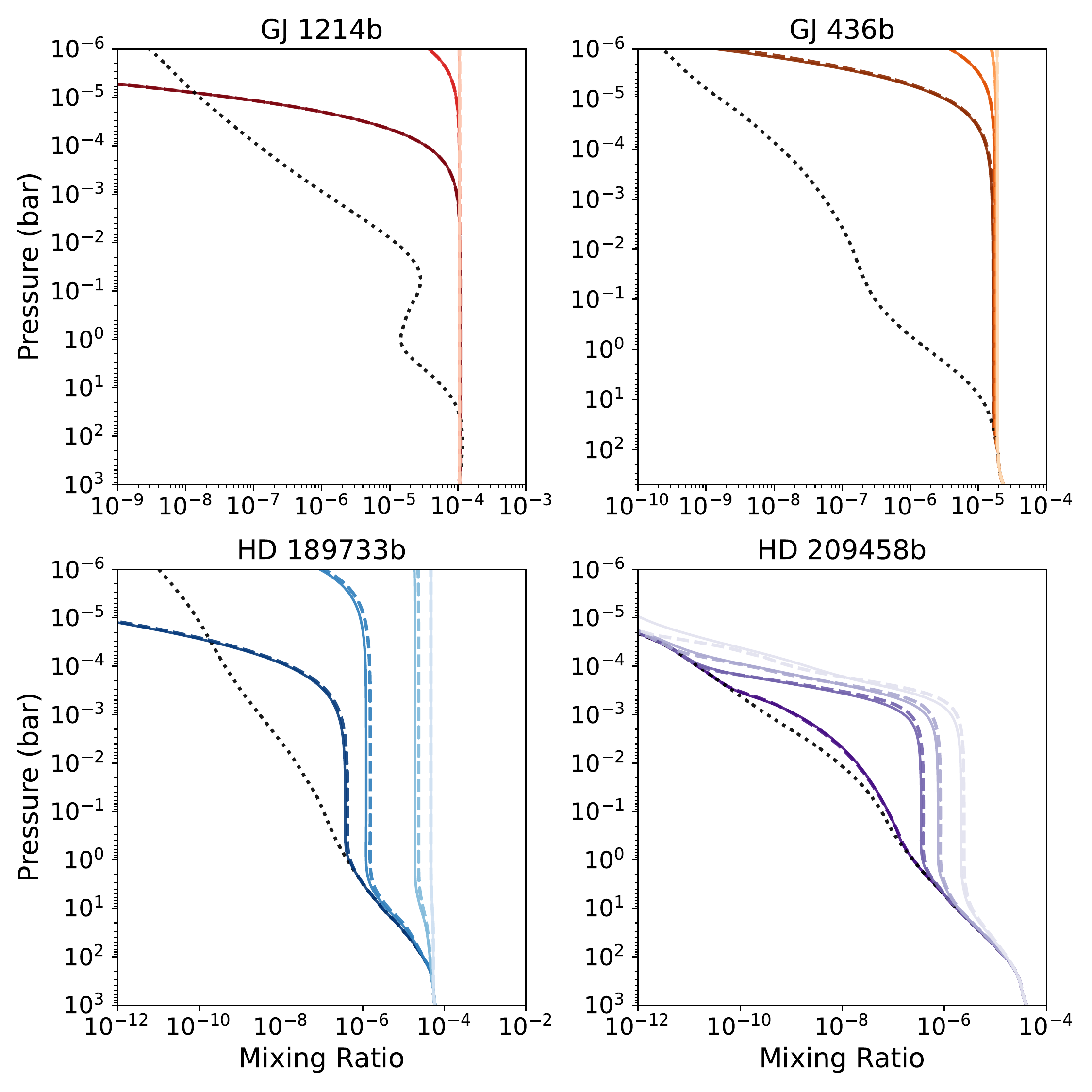}
  \subcaption{\ce{NH3}}
  \end{subfigure}
    \begin{subfigure}[t]{0.475\linewidth}
  \includegraphics[width=\columnwidth]{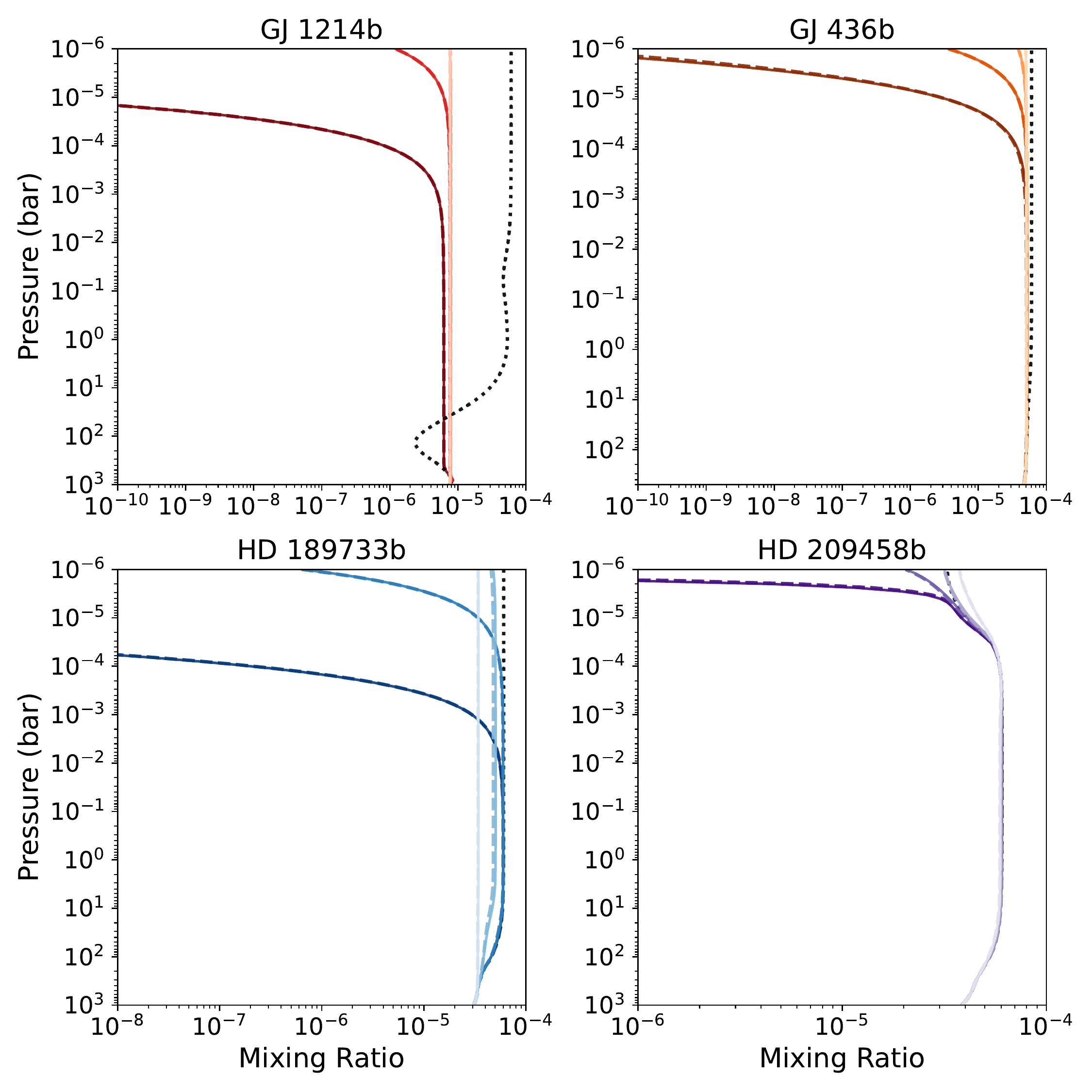}
  \subcaption{\ce{N2}}
  \end{subfigure}
  \caption{(cont.)}
  \end{figure*}

% \begin{figure}\label{-}
%   \centering
%   %\includegraphics[width=\hsize]{figs/H2O-conden}
%     \caption{Quenching validation (3a): H, CH4, CO, CO2, C2H2, C2H6, HCN, NH3 for one TP and various Kzz (other TPs in appendix) (3b): for 500X metallicity (3c): for C/O =  2}
%   \end{figure}

% \begin{figure}\label{-}
%   \centering
%   %\includegraphics[width=\hsize]{figs/H2O-conden}
%     \caption{Photochemical validation with the same parameters.\\}
%   \end{figure}

\begin{table*}[!h]
\begin{center}
\caption{A selected portion of the table of the rate constants for the net reaction \ce{CH4 + H2O -> CO + 3H2}.}
\begin{tabular}{llll}
% \hline
% \hline
Temperature (K) & Pressure (bar) & Rate Constant (cm$^3$ molecules$^{-1}$ s$^{-1}$) & Rate-limiting Step\\
\hline
300 & 10$^{-6}$ & 1.332 $\times$ 10$^{-80}$& \ce{CH3OH + H -> CH3 + H2O}\\
$\vdots$ & $\vdots$ & $\vdots$ & $\vdots$\\
1000 & 10$^{-1}$ & 9.244 $\times$ 10$^{-30}$ &  \ce{CH2OH + H -> OH + CH3}\\
$\vdots$ & $\vdots$ & $\vdots$ & $\vdots$\\
1000 & 7.943 $\times$ 10$^{-1}$  & 4.924 $\times$ 10$^{-30}$ &  \ce{OH + CH3 ->[M] CH3OH}\\
$\vdots$ & $\vdots$ & $\vdots$ & $\vdots$\\
2000 & 1 & 7.005 $\times$ 10$^{-17}$ & \ce{CH2OH + ->[M] H + H2CO}\\
$\vdots$ & $\vdots$ & $\vdots$ & $\vdots$\\
\hline
\end{tabular}\label{tab:ex-rates}
\end{center}
\end{table*}

\section{Results}
\subsection{0D validation: the same chemical timescales produced even with different paths}\label{sec:0D} %  at times
Fig. \ref{0D-evo} compares the temporal evolution of the major species computed by the mini-networks and full chemical kinetics (VULCAN). In most cases, the mini-network manifests the same temporal path as the full kinetics. At high temperatures and low pressures ($T \gtrsim 1500$ K and $P \lesssim$ 1 mbar), the evolution from the mini-network can start to take somewhat different paths (e.g., \ce{CH4} and \ce{NH3} in the upper right panel of Fig. \ref{0D-evo}). This is likely due to more participation of small molecules and atoms in this regime, such as C and CN, that are not included in the mini-network. Despite different paths, these species still achieve the equilibrium state around the same time as those in the full kinetics. Defining the timescale as the time it takes for the composition to approach the equilibrium value within 0.1 $\%$, we evaluate the relative errors (|t$_{\textrm{net}}$-t$_{\textrm{full}}$|/t$_{\textrm{full}}$ $\times$ 100$\%$) of the chemical timescales for the main species (\ce{H2O}, \ce{CH4}, CO, \ce{CO2}, \ce{NH3}, HCN, and \ce{N2}) in solar metallicity. We find the maximum relative error of 135 $\%$ and the mean relative error of 43 $\%$ in the most relevant range of 1000 K $\leq T \leq$ 2500 K and 10$^{-4}$ bar $\leq P \leq$ 1000 bar. The 0D tests show that the mini-network scheme can successfully reproduce the chemical timescale from the full kinetics.     

% {\bf $\sim$ 20 times faster than the full kinetics in VULCAN}\\

%  at various temperatures and pressures
% C produced faster in full kinetics for T=2000K, 1mbar, at t=1 sec is due to \ce{H + CH -> H2 + C}\\

\subsection{1D validation: reproducing correct quench levels}
The vertical distributions of the main compositions computed by the two chemical networks of each planet with different vertical mixing for solar metallicity are summarized in Fig. \ref{fig:1D}. Among the explored eddy diffusion coefficients from 10$^5$ to 10$^{11}$ cm$^2$/s, we present the resulting profiles that are sensitive to the change for clarity. Specifically, $K_{\textrm{zz}}$ = 10$^5$, 10$^7$, 10$^9$, 10$^{11}$ (cm$^2$/s) are shown for GJ 1214b, $K_{\textrm{zz}}$ = 10$^6$, 10$^7$, 10$^8$, 10$^9$  (cm$^2$/s) for GJ 436b, $K_{\textrm{zz}}$ = 10$^5$, 10$^7$, 10$^9$, 10$^{11}$ (cm$^2$/s) for HD 189733b, and $K_{\textrm{zz}}$ = 10$^7$, 10$^9$, 10$^{10}$, 10$^{11}$ (cm$^2$/s) for HD 209458b. First of all, transport-induced quenching is correctly reproduced by the mini-network, i.e., the quench levels of CO on the cooler planets GJ 1214b and GJ 436b and those of \ce{CH4} on the hotter planets HD 189733b and HD 209458b agree well between two networks. For species that react fast with the major species, such as \ce{C2H2} and HCN, the non-constant mixing ratio profiles as they follow their parent molecules before themselves quenched at lower pressure \citep[e.g.,][]{Moses11,tsai17} are also well captured by the mini-network.

Of all species in the mini-network, \ce{C2H2} appears to have the largest deviation, up to about a factor of three on GJ 1214b, which can be attributed to the combination of its low abundance and the simplification of hydrocarbon kinetics. The mini-network is able to correctly reproduce the vertical quenching of the main species and notably the abundance profiles with the second equilibrium region present in the upper atmosphere due to thermal inversion on HD 209458b. The maximum errors\footnote{|x$_{\textrm{net}}$-x$_{\textrm{full}}$|/x$_{\textrm{full}}$ $\times$ 100 $\%$} of the main species with mixing ratios not lower than 10$^{-20}$ in the region of observational interest (1 bar -- 0.1 mbar) computed by the mini-chemical scheme for the whole ranges of eddy diffusion coefficients are listed in Table \ref{tab:err}. Those errors greater than 100 $\%$ all occur with volume mixing ratios smaller than 10$^{-10}$. We find the discrepancies for the main species abundances between the mini-network and the full kinetics always less than an order of magnitude and rarely exceeds a factor of two, consistent with the 0D validation in Sect. \ref{sec:0D}.

A self-contained way to evaluate the errors with respect to the presumed uncertainty factors in the reaction rates of the full kinetics is to perform an uncertainty propagation analysis \citep{dobrijevic10,Wakelam2010}. \cite{Venot2019} determined a tighter constraint of 10$\%$ for their GJ 436b model
by the Monte-Carlo uncertainty propagation \citep{Hebard2015} with the uncertainty factor derived by the combustion study. However, in practice, it is not uncommon to have models with different sets of kinetics data differ by an order of magnitude \citep{Moses2014,Tsai2021} when the overall aspects of uncertainties are taken into account. Therefore, at least before the kinetics discrepancies are fully resolved, we consider an order of magnitude of error to be acceptable for exoplanet application.

% {\bf Overall, we find that discrepancies for the main species abundances between the mini-network and the full kinetics are always less than an order of magnitude and rarely exceeds a factor of two, consistent with the 0D validation in Sect. \ref{sec:0D}.} 

% We now validate the mini-chemical scheme with the composition profiles computed based on the  using the 1D temperature profiles in Fig. \ref{TPs}. 

% Describe Kzz. The results that are sensitive to the change of eddy diffusion are shown (lighter colors correspond to larger $K_{\textrm{zz}}$): $K_{\textrm{zz}}$ = 10$^5$, 10$^7$, 10$^9$, 10$^11$ cm$^2$/s for 51 Eri b; 10$^6$, 10$^7$, 10$^8$, 10$^9$  cm$^2$/s for GJ 436b; 10$^5$, 10$^7$, 10$^9$, 10$^11$ cm$^2$/s for HD 189733b; 10$^7$, 10$^9$, 10$^10$, 10$^11$ cm$^2$/s for HD 209458b. We find the differences between the mini-network and the full kinetics are generally within a factor of two, consistent with the validation in \ref{sec:0D}.
\begin{table}[htp]
\begin{center}
\caption{The maximum errors in $\%$ of the compositions in the pressure range of 1 bar -- 0.1 mbar computed by the mini-network for solar elemental abundances, with the largest errors among four planets shown in bold. The compositions obtained by the full kinetics from VULCAN serves as the reference.}
%(backward reactions with even indexes are reversed numerically with thermodynamic data) and rate coefficients (cm$^3$ s$^{-1}$ for bimolecular reactions and s$^{-1}$ for k$_0$) of }
%\tiny
\begin{tabular}{l|l|l|l|l}
%\hline
Species & GJ 1214b & GJ 436b & HD 189733b & HD 209458b\\
\hline
\ce{H2O} & 0.07 & 0.0002 & {\bf 15} & 2\\
\ce{CH4} & 0.07 & 0.001 & 22 & {\bf 144}\\
\ce{CO} & 4 & 3 & {\bf 12} & 1\\
\ce{CO2} & 4 & 2 & {\bf 6} & 0.9\\
\ce{C2H2} & {\bf 284} & 137 & 89 & 141\\
\ce{NH3} & 0.07 & 0.02 & 21 & {\bf 90}\\
\ce{N2} & 0.1 & 0.1 & {\bf 5} & 0.4\\
\ce{HCN} & 13 & 13 & 36 & {\bf 155}\\
\end{tabular}\label{tab:err}
\end{center}
\end{table}

\subsection{Varying metallicity and C/O ratio}
%discuss significant change in the pathways such as C2H2?\\
While the rate coefficients of net reactions depend on the elemental abundances, the approach is general in principle. The same procedure may be applied to construct a mini-network with relevant elementary and net reactions for arbitrary elemental composition. Here, we vary the metallicity and C/O ratio to test the validity of our mini-network designed for \ce{H2}-dominated composition. First, Figures \ref{fig:rates_T2000} illustrate how the rate coefficients of \ce{CH4 + H2O -> CO + 3H2} and \ce{2NH3 -> N2 + 3H2} vary with metallicity and C/O ratio. The rate coefficients of \ce{CH4} $\rightarrow$ \ce{CO} and \ce{NH3} $\rightarrow$ \ce{N2} generally increase with metallicity, while no consistent trends are found for the C/O ratio. Next, Figures \ref{fig:0D-evo-100X} -- \ref{fig:1D-solarCtoO2} of Appendix \ref{app} showcase the same 0D and 1D validation with 500 times solar metallicity and solar composition but with C/O = 2.The maximum errors are also listed in Tables \ref{tab:err-500} and \ref{tab:err-ctoo2}. We find the scheme less accurate with higher metallicity when the atmosphere becomes less \ce{H2}-dominated. Therefore, we restrict the valid range of our mini-network to not exceeding 500 times solar metallicity. Of all the explored cases, \ce{C2H2} remains associated with the largest error, and less abundant species tend to have larger errors too. For instance, \ce{CH4} and \ce{NH3} produce bigger errors in hotter planets, HD 189733b and HD 209458b, whereas \ce{C2H2} is more accurate in warm conditions where it is favored. Compared to the updated reduced chemical scheme (with a new \ce{CH3OH} mechanism and including \ce{C2H2}) in \cite{Venot2020}, the mini-network achieves comparable accuracy ($\lesssim$ 10 $\%$) for GJ 436b with a solar metallicity except for \ce{C2H2}. The same trend of increased errors with higher metallicity is also found in the reduced chemical scheme of \cite{Venot2020}. For the hot Jupiters HD 189733b and HD 209458b, the mini-network produces more significant errors in \ce{C2H2} while the reduced network in \cite{Venot2020} appears to produce larger errors in \ce{NH3}. Overall, the agreement between our mini-chemical scheme and the full kinetics remains well under an order of magnitude, similar to the accuracy of the reduced scheme (with 44 species and 582 reactions) from \cite{Venot2020}.\\

\section{Conclusions} % 500 K $\leq T \leq$ 3000 K and 10$^{-6}$ bar $\leq P \leq$ 10$^3$ bar
We have devised a novel chemical scheme utilizing net reactions to significantly reduce the size of a chemical network. The new scheme is validated across a wide range of temperatures and pressures by comparing the chemical timescales from the mini-network and the full kinetics VULCAN \citep{tsai17,Tsai2021}. The mini-network scheme is able to reproduce the quenching behavior of major species well under an order of magnitude in the benchmark exoplanet atmospheres (GJ 1214b, GJ 436b, HD 189733b, HD 209458b). The tabulated rates of the net reactions from 300 K $\leq T \leq$ 3000 K, 10$^{-6}$ bar $\leq P \leq$ 10$^3$ bar for the valid ranges of metallicities (0.1 -- 500 times solar) and C/O (0.25 -- 2 times solar) are available in the supplementary files. The presented scheme is robust yet simple to adopt and fast to run. The mini-network takes about 1.5 $\times$ 10$^{-3}$ s (tested on a 2015 laptop with 2.2 GHz Intel Core i7 using SciPy linear algebra routines) to integrate an atmospheric cell for one time step. For comparison, its computational time is about 25 times faster than the original C-H-N-O network in \cite{Tsai2021b} and about 10 times faster than a network with a size similar to that of \cite{Venot2019}\footnote{We performed the test with a C-H-O network of 34 species and 362 total reactions}. We hope it will encourage the field of research moving forward to incorporate a more realistic chemical mechanism in 3D models and retrieval frameworks.  

We reiterate that unlike the relaxation method \citep{Cooper2006,Drummond2018,tsai18}, the mini-chemical network keeps the same form of rate equations as the standard kinetics. In addition to the major molecules of observational interest or radiative importance, key radical species are also included. This allows us to extend the scheme to incorporate photochemistry, where radical species are produced by photodissociation. We will present the detailed treatment of photochemistry in a follow-up paper.

%provided for T from 300 to 3000 K, solar (10x, 100x, 1000x) metallicity, C/O (1, 1.5, 2?) ... etc. \\

%   \begin{enumerate}
%       \item A
%       \item B
%       \item C
%   \end{enumerate}
% Tables with RLS are availabe onine at ...\\
\begin{acknowledgements} 
S.-M. T. thanks F. Selsis for the project collaboration that sparks the mini-network conception and T. Fisher for comments on the graph analysis. S.-M. T. acknowledges support from the European community through the ERC advanced grant EXOCONDENSE (\#740963; PI: R.T. Pierrehumbert).
\end{acknowledgements}

\bibliographystyle{aa}
\bibliography{master_bib}

% WARNING
%-------------------------------------------------------------------
% Please note that we have included the references to the file aa.dem in
% order to compile it, but we ask you to:
%
% - use BibTeX with the regular commands:
%   \bibliographystyle{aa} % style aa.bst
%   \bibliography{Yourfile} % your references Yourfile.bib
%
% - join the .bib files when you upload your source files
%-------------------------------------------------------------------

\begin{appendix}
%\nopartblankpage
\section{Validation for nonsolar elemental abundances}\label{app}
%{\bf (Remove this blank page)}

\begin{figure*}[!htp]
\centering
   \includegraphics[width=0.49\hsize]{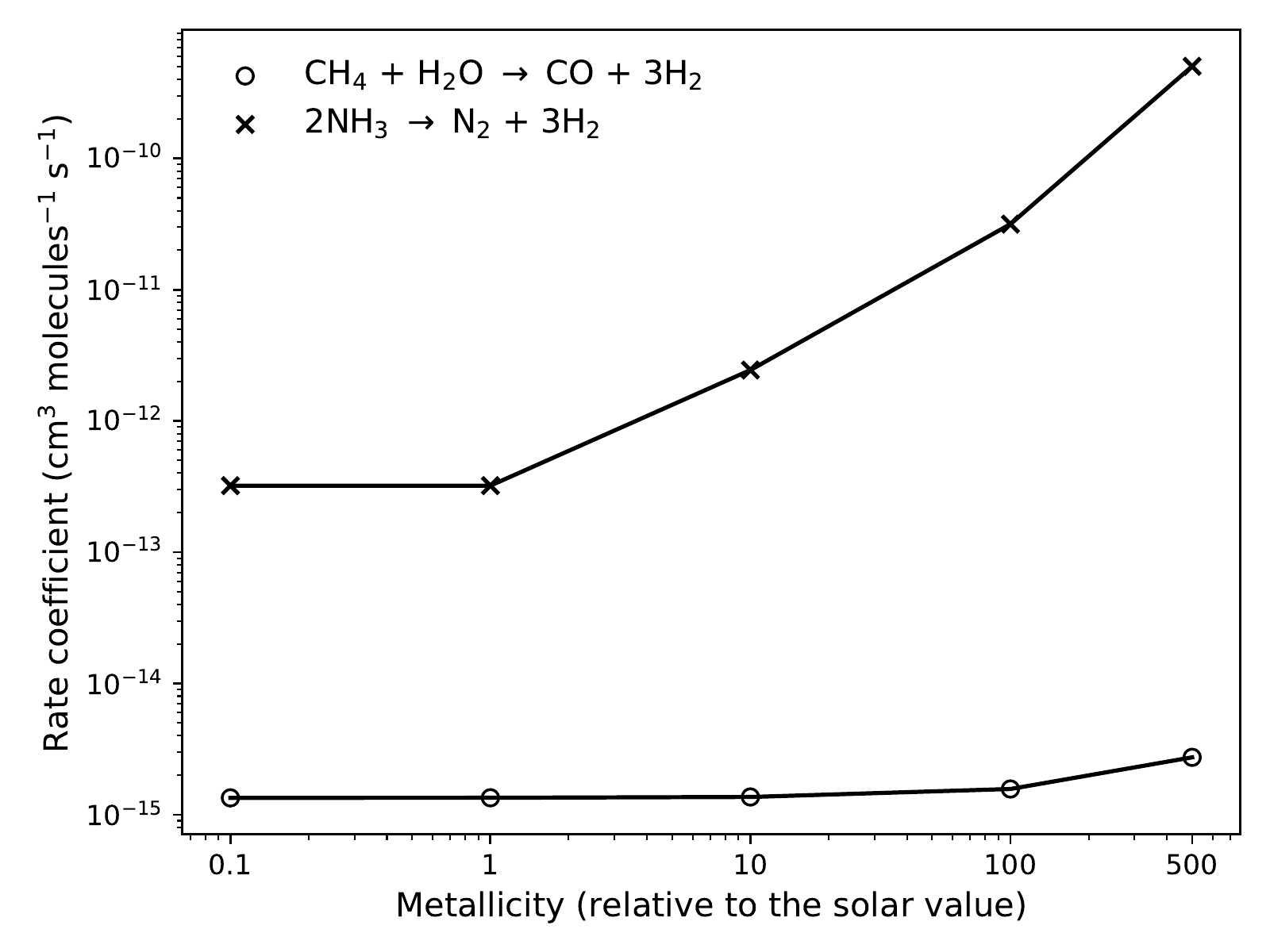}
   \includegraphics[width=0.49\hsize]{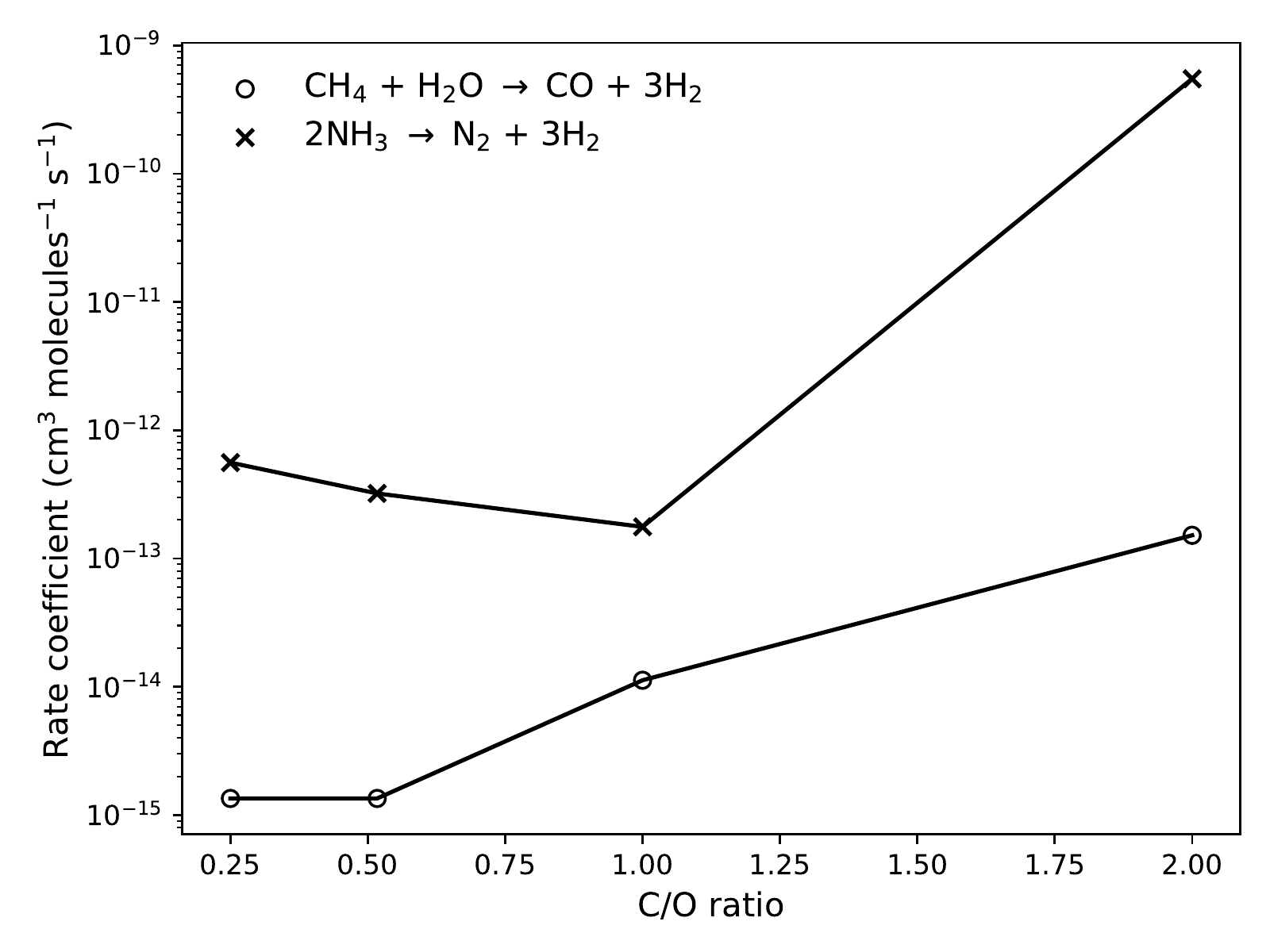}
    \caption{The rate coefficients of the net reactions \ce{CH4 + H2O -> CO + 3H2} and \ce{2NH3 -> N2 + 3H2} at 2000 K and 0.01 bar as a function of metallicity (left) and C/O ratio (right). }\label{fig:rates_T2000}
\end{figure*}

\begin{figure*}[!h]
   \centering
   \includegraphics[width=\hsize]{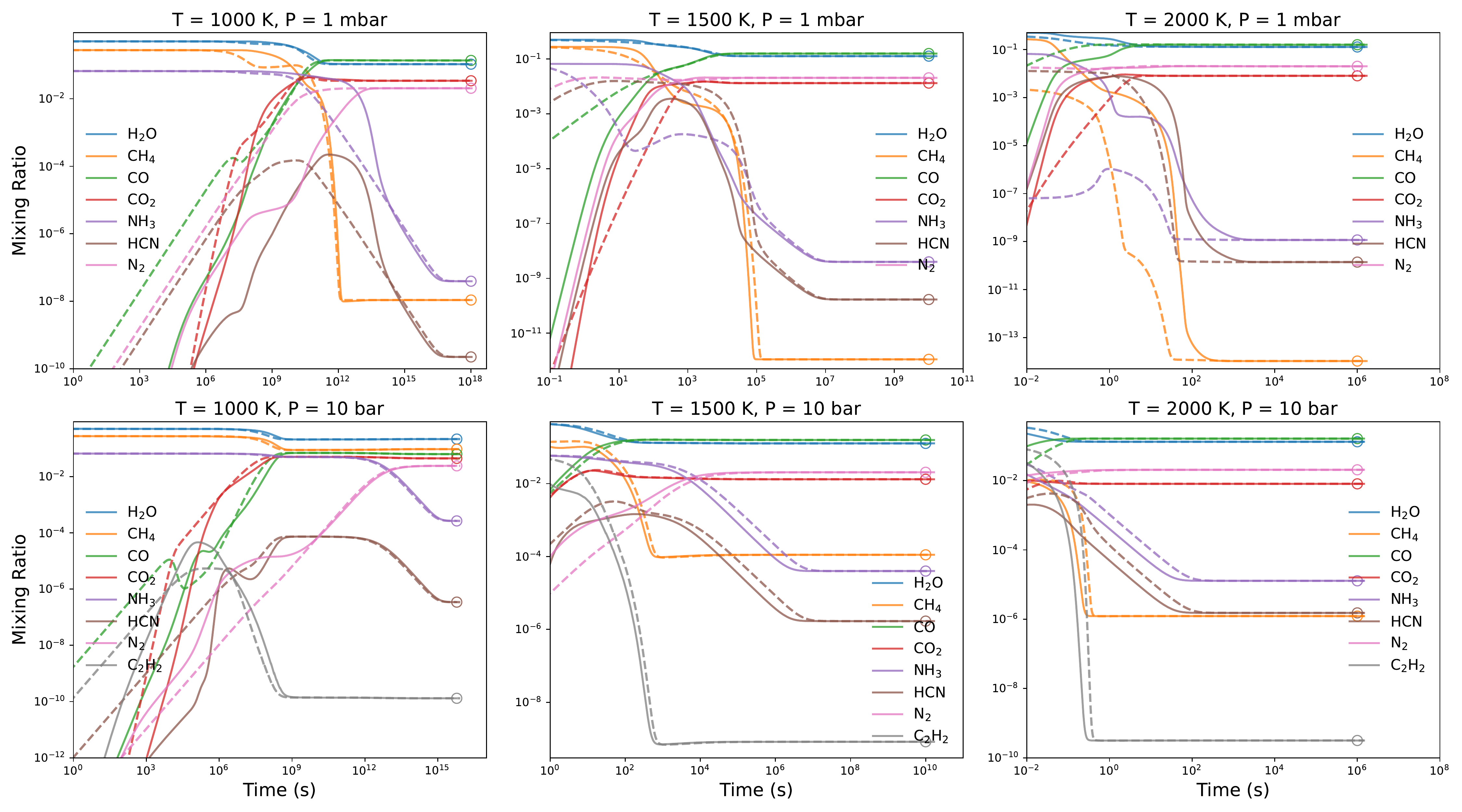}
    \caption{Same as Fig. \ref{0D-evo} but for 500 $\times$ solar metallicity}\label{fig:0D-evo-100X}
   \end{figure*}
\begin{figure*}[!h]
   \centering
   \includegraphics[width=\hsize]{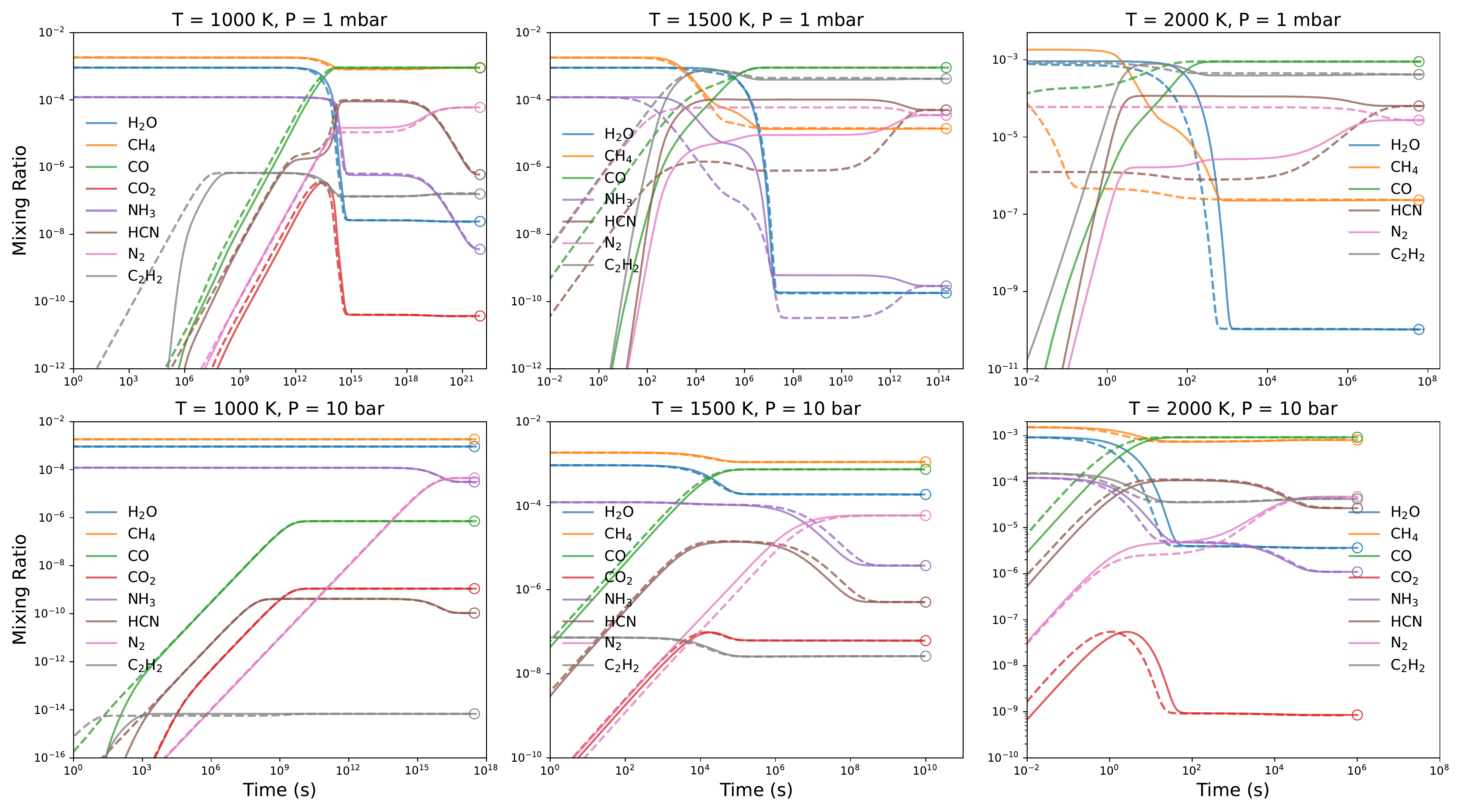}
    \caption{Same as Fig. \ref{0D-evo} but for solar metallicity except for C raised to C/O = 2}\label{fig:0D-evo-solarCtoO2}
   \end{figure*}
%{\bf 1D results: HCN on 51 Eri b is not right?}   

\begin{table}[!hp]
\begin{center}
\caption{Same as Table \ref{tab:err} but for 500$\times$ solar metallicity.}
%(backward reactions with even indexes are reversed numerically with thermodynamic data) and rate coefficients (cm$^3$ s$^{-1}$ for bimolecular reactions and s$^{-1}$ for k$_0$) of }
%\tiny
\begin{tabular}{l|l|l|l|l}
%\hline
Species & GJ 1214b & GJ 436b & HD 189733b & HD 209458b\\
\hline
\ce{H2O} & 2 & 0.4 & {\bf 10} & 2\\
\ce{CH4} & 2 & 0.6 & 30 & {\bf 84}\\
\ce{CO} & 8 & 2 & {\bf 15} & 2\\
\ce{CO2} & 2 & 2 & {\bf 25} & 1\\
\ce{C2H2} & {\bf 972} & {\bf 972} & 52 & 887\\
\ce{NH3} & 8 & 3 & 53 & {\bf 134}\\
\ce{N2} & 0.8 & 0.2 & {\bf 16} & 2\\
\ce{HCN} & 72 & 39 & 69 & {\bf 916}\\
\end{tabular}\label{tab:err-500}
\end{center}
\end{table}

\begin{table}[!hp]
\begin{center}
\caption{Same as Table \ref{tab:err} but for C raised to C/O = 2.}
%(backward reactions with even indexes are reversed numerically with thermodynamic data) and rate coefficients (cm$^3$ s$^{-1}$ for bimolecular reactions and s$^{-1}$ for k$_0$) of }
%\tiny
\begin{tabular}{l|l|l|l|l}
%\hline
Species & GJ 1214b & GJ 436b & HD 189733b & HD 209458b\\
\hline
\ce{H2O} & 0.001 & 0.0005 & 55 & {\bf 238}\\
\ce{CH4} & 0.002 & 0.002 & 7 & {\bf 269}\\
\ce{CO} & 10 & 2 & {\bf 12} & 0.1\\
\ce{CO2} & 10 & 3 & 36 & {\bf 238}\\
\ce{C2H2} & {\bf 367} & 151 & 103 & 19\\
\ce{NH3} & 0.7 & 0.06 & 10 & {\bf 90}\\
\ce{N2} & 5 & 0.3 & {\bf 12} & 5\\
\ce{HCN} & 13 & 13 & 12 & {\bf 25}\\
\end{tabular}\label{tab:err-ctoo2}
\end{center}
\end{table}

\begin{figure*}
  \centering
  \begin{subfigure}[t]{0.475\linewidth}
  \includegraphics[width=\columnwidth]{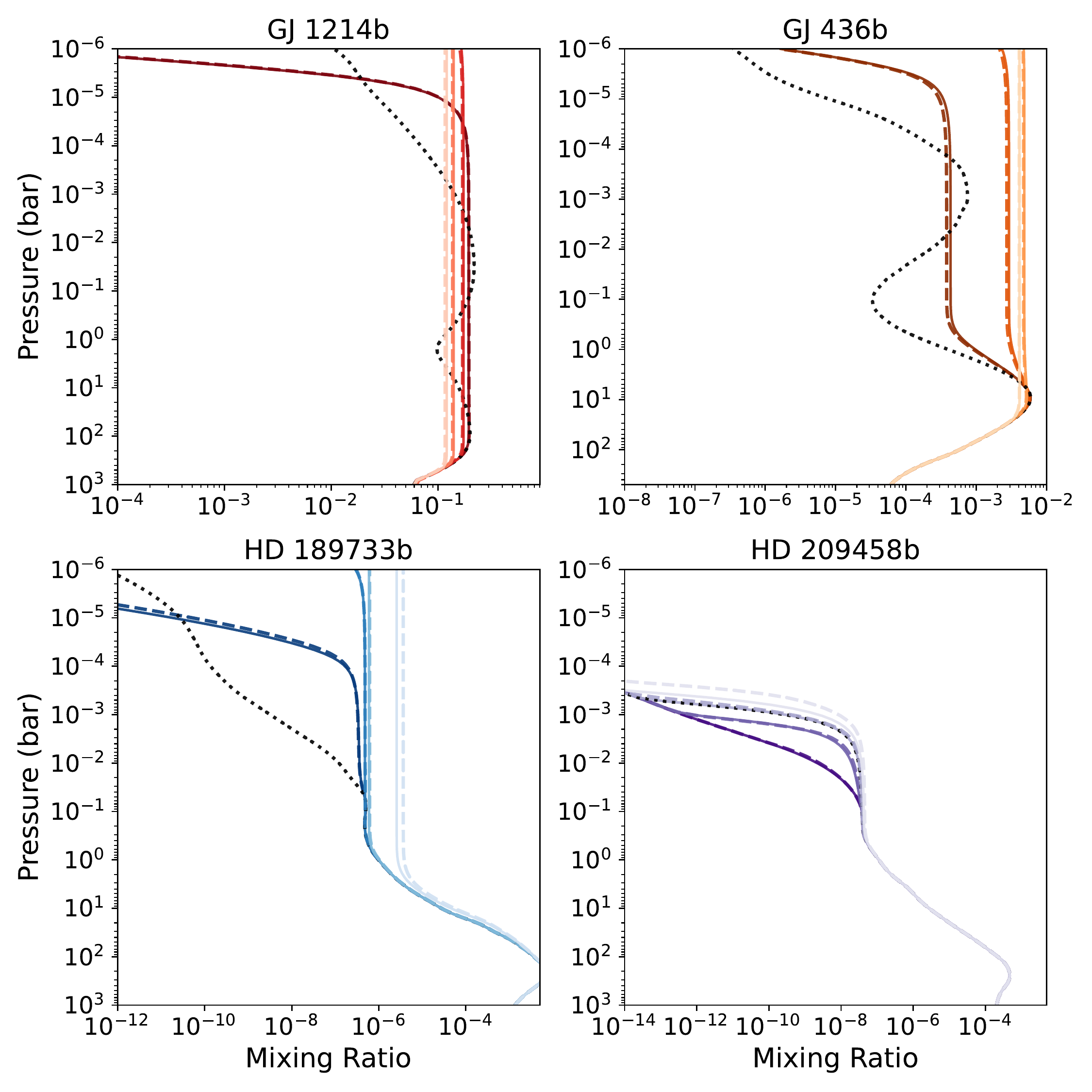}
  \subcaption{\ce{CH4}}
  \end{subfigure}
  \begin{subfigure}[t]{0.475\linewidth}
  \includegraphics[width=\columnwidth]{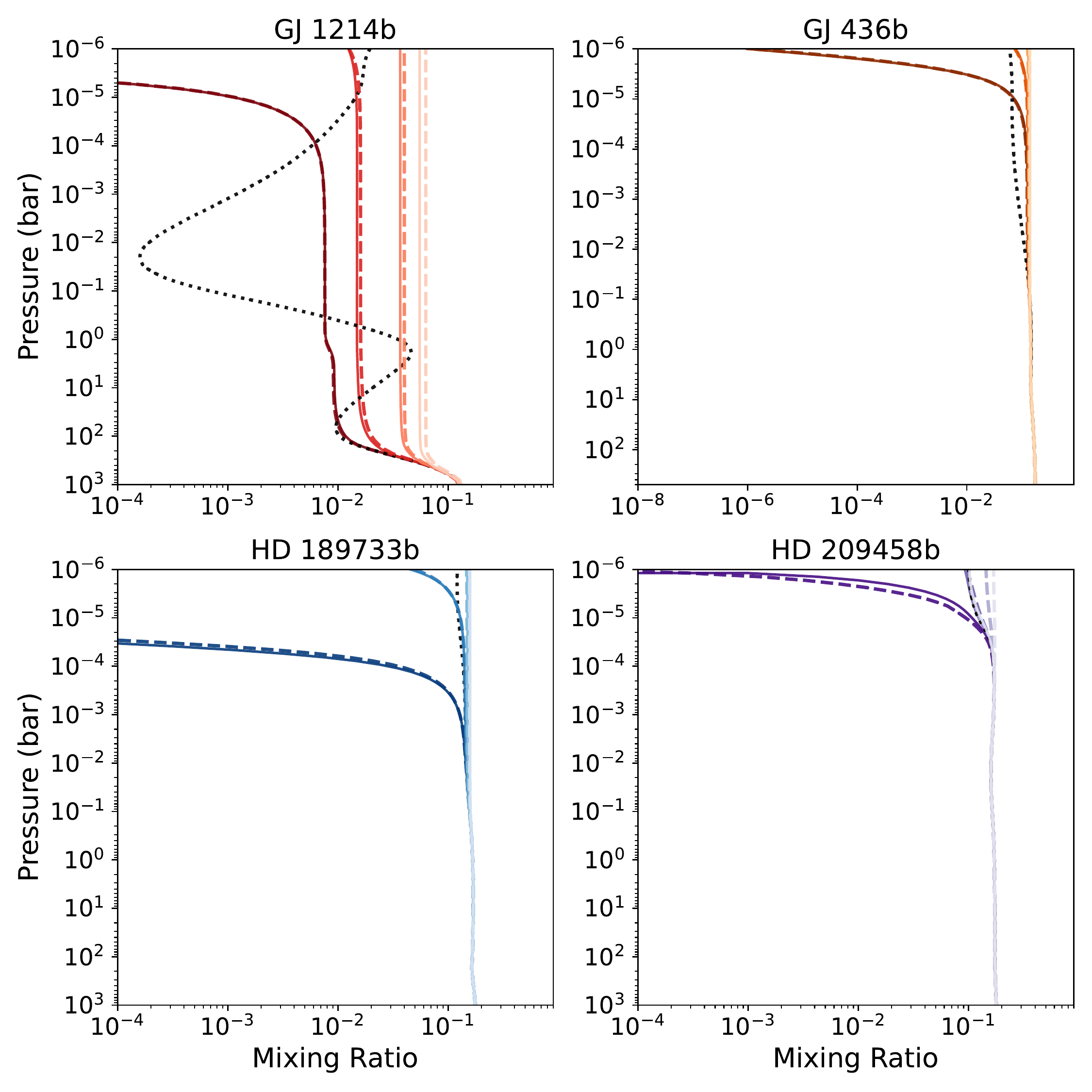}
  \subcaption{\ce{CO}}
  \end{subfigure}
  \begin{subfigure}[t]{0.475\linewidth}
  \includegraphics[width=\columnwidth]{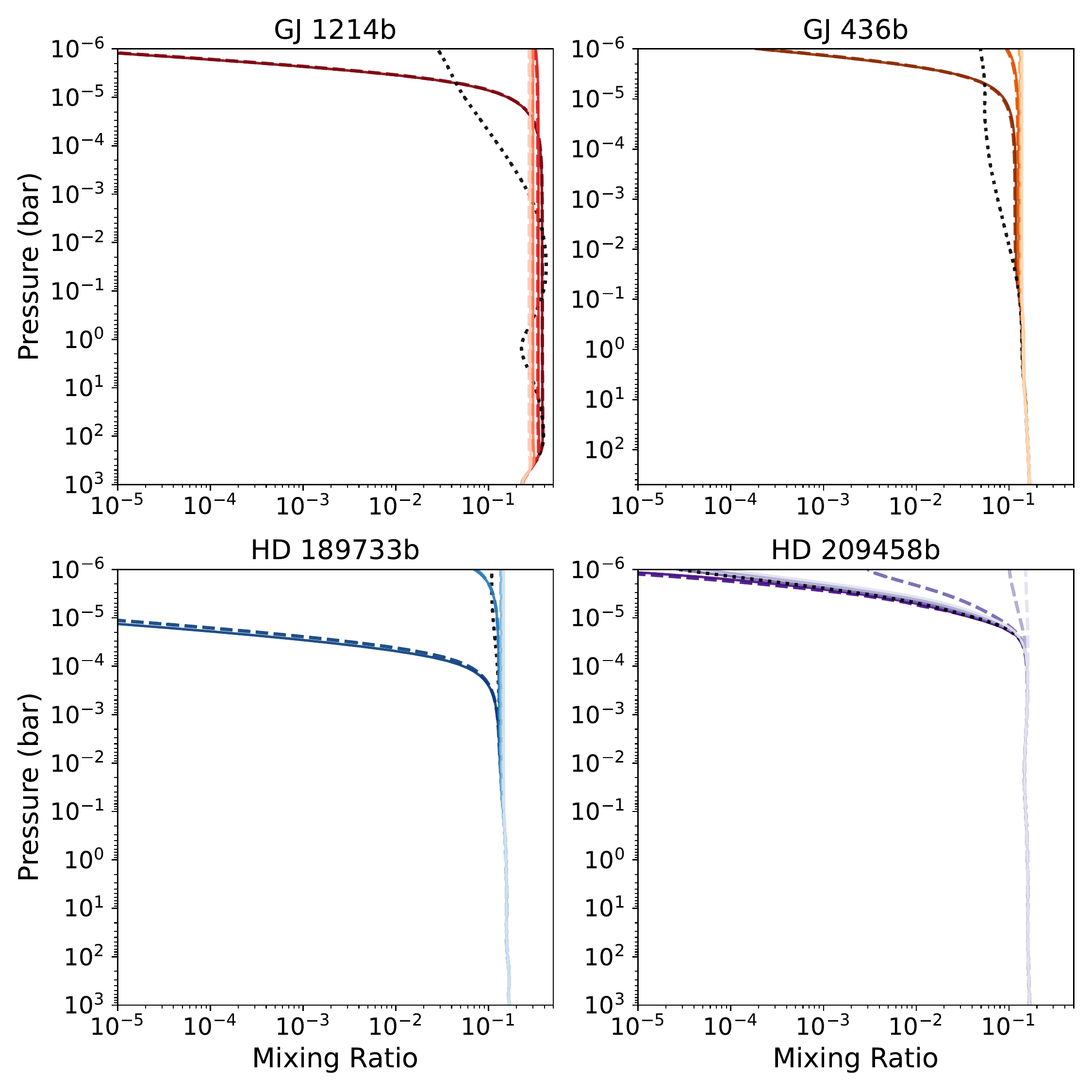}
  \subcaption{\ce{H2O}}
  \end{subfigure}
    \begin{subfigure}[t]{0.475\linewidth}
  \includegraphics[width=\columnwidth]{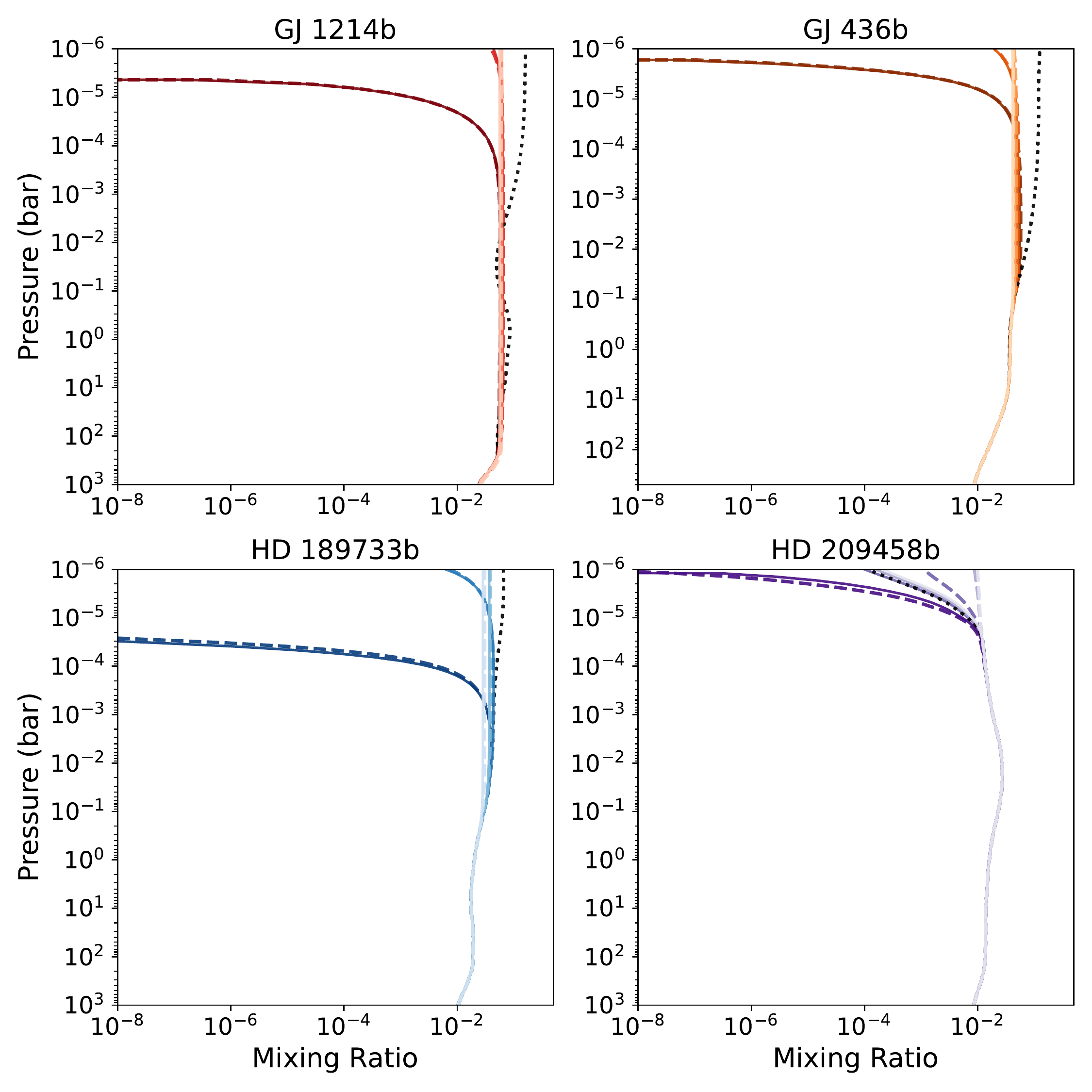}
  \subcaption{\ce{CO2}}
  \end{subfigure}
  \caption{Same as Fig. \ref{fig:1D} but for 500 $\times$ solar metallicity.}\label{fig:1D-100Xsolar} % Not showing the compositions with negligible mixing ratios ( $<$ 10$^{-12}$ between 1 and 10$^{-5}$ bar)
  \end{figure*}

\begin{figure*}
\ContinuedFloat
  \centering
  \begin{subfigure}[t]{0.475\linewidth}
  \includegraphics[width=\columnwidth]{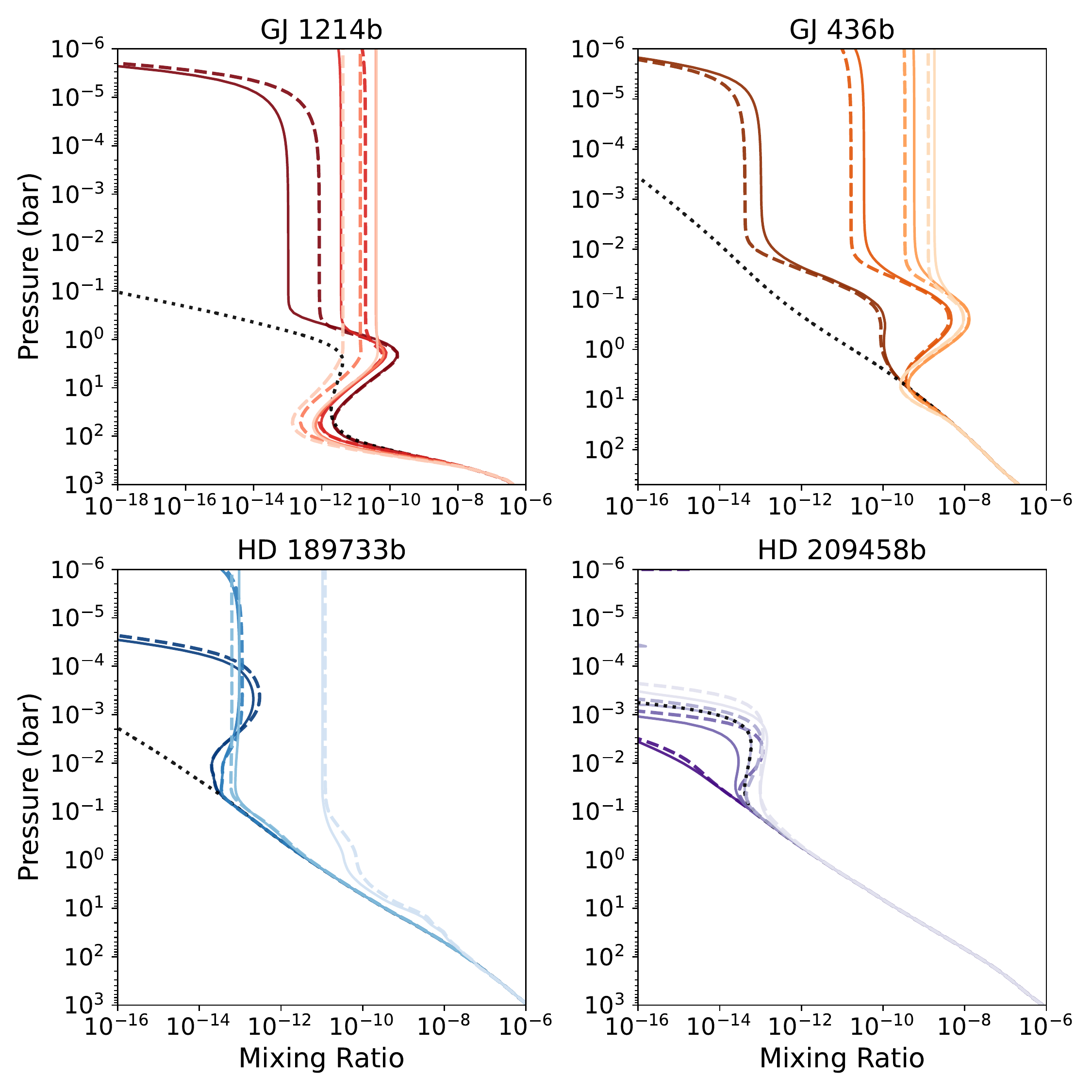}
  \subcaption{\ce{C2H2}}
  \end{subfigure}
  \begin{subfigure}[t]{0.475\linewidth}
  \includegraphics[width=\columnwidth]{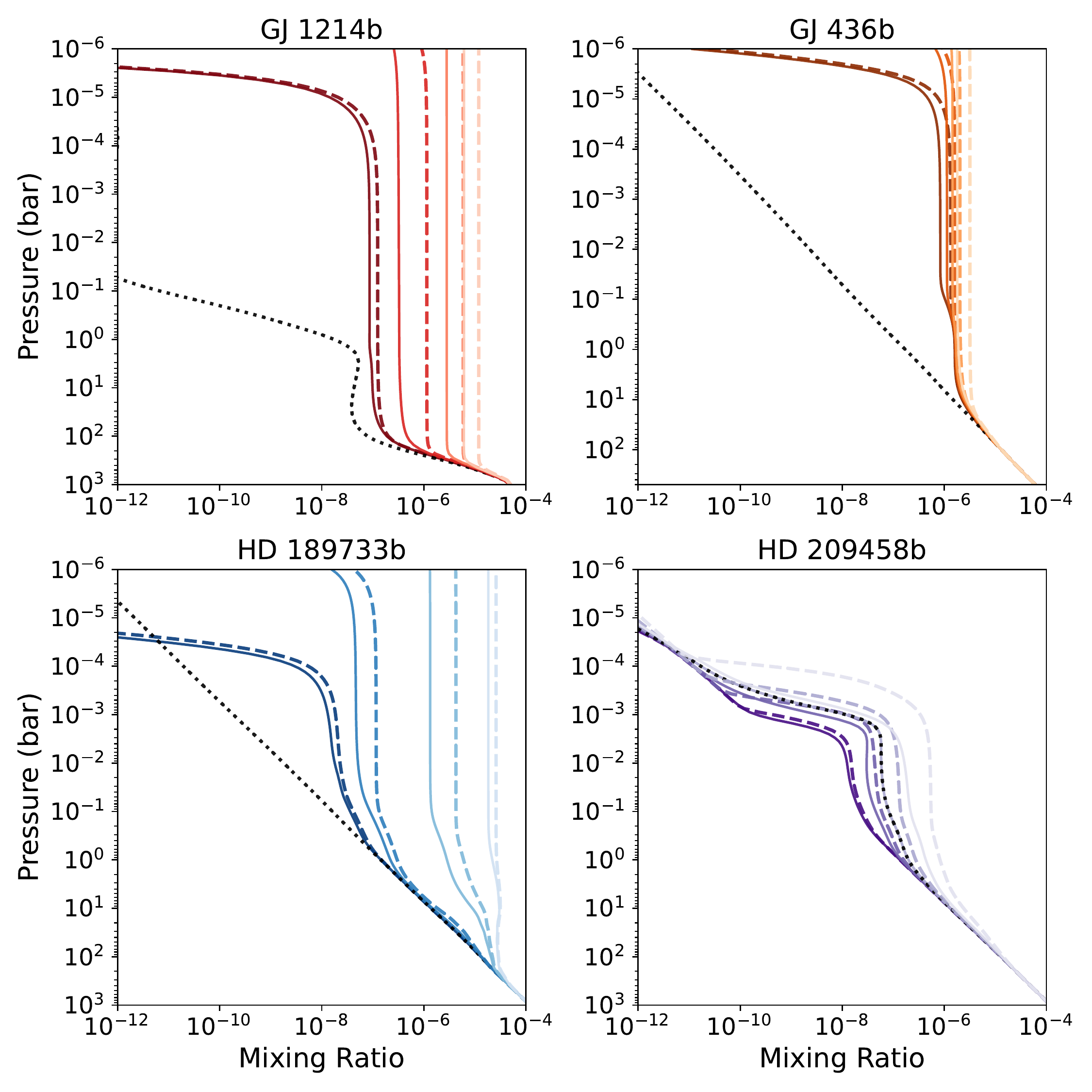}
  \subcaption{\ce{HCN}}
  \end{subfigure}
  \begin{subfigure}[t]{0.475\linewidth}
  \includegraphics[width=\columnwidth]{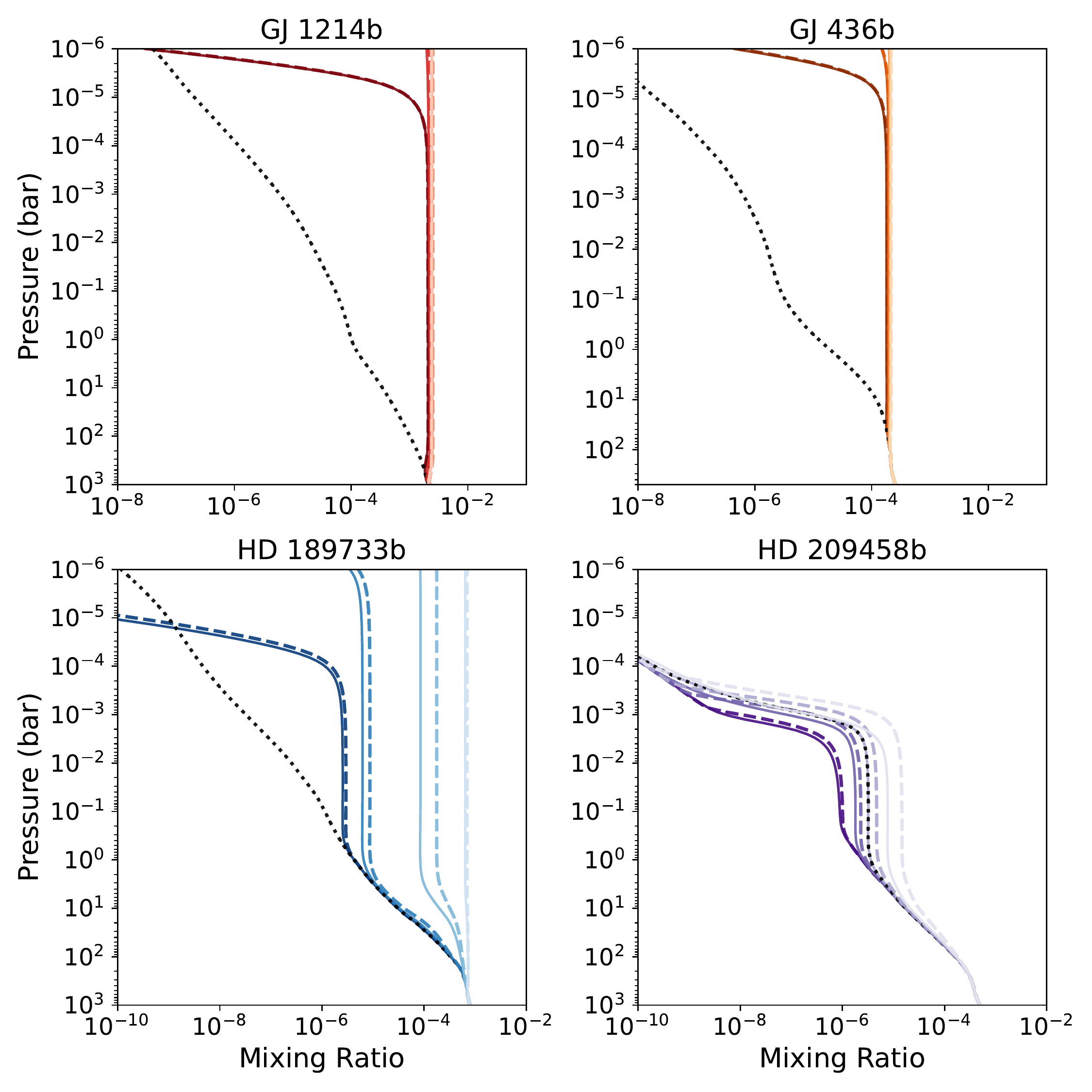}
  \subcaption{\ce{NH3}}
  \end{subfigure}
    \begin{subfigure}[t]{0.475\linewidth}
  \includegraphics[width=\columnwidth]{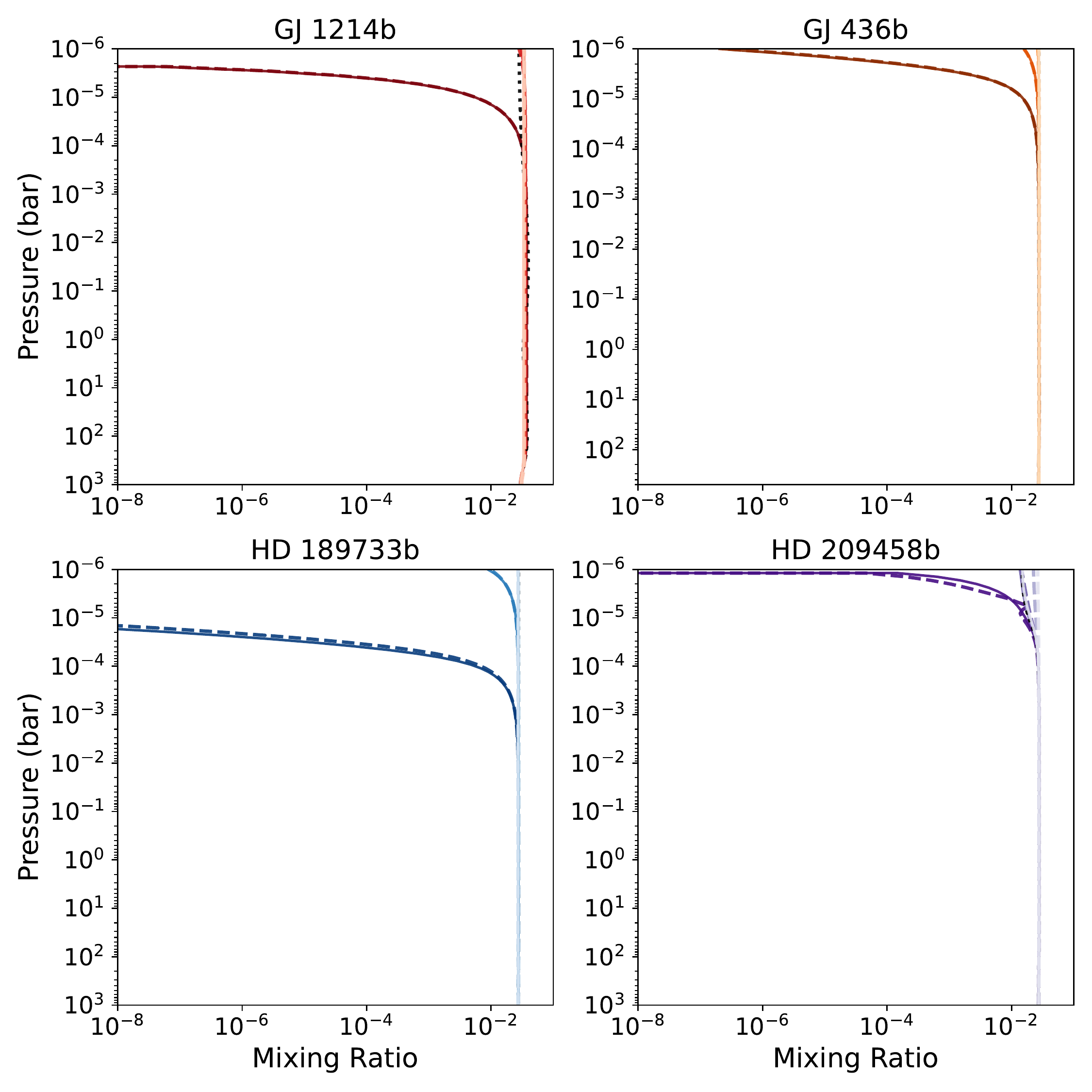}
  \subcaption{\ce{N2}}
  \end{subfigure}
  \caption{(cont.)}
  \end{figure*}

\begin{figure*}
  \centering
  \begin{subfigure}[t]{0.475\linewidth}
      \includegraphics[width=\columnwidth]{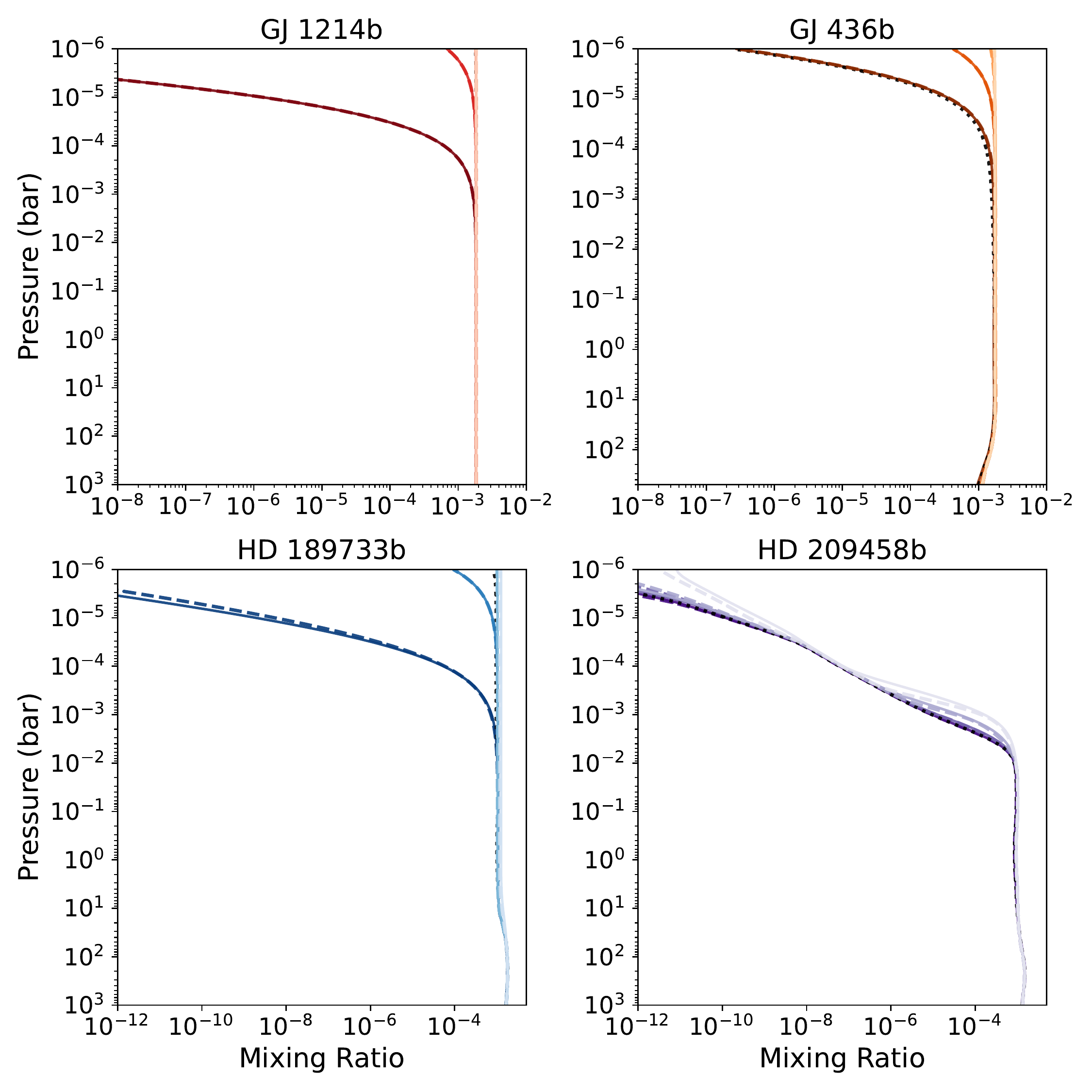}
  \subcaption{\ce{CH4}}
  \end{subfigure}
  \begin{subfigure}[t]{0.475\linewidth}
  \includegraphics[width=\columnwidth]{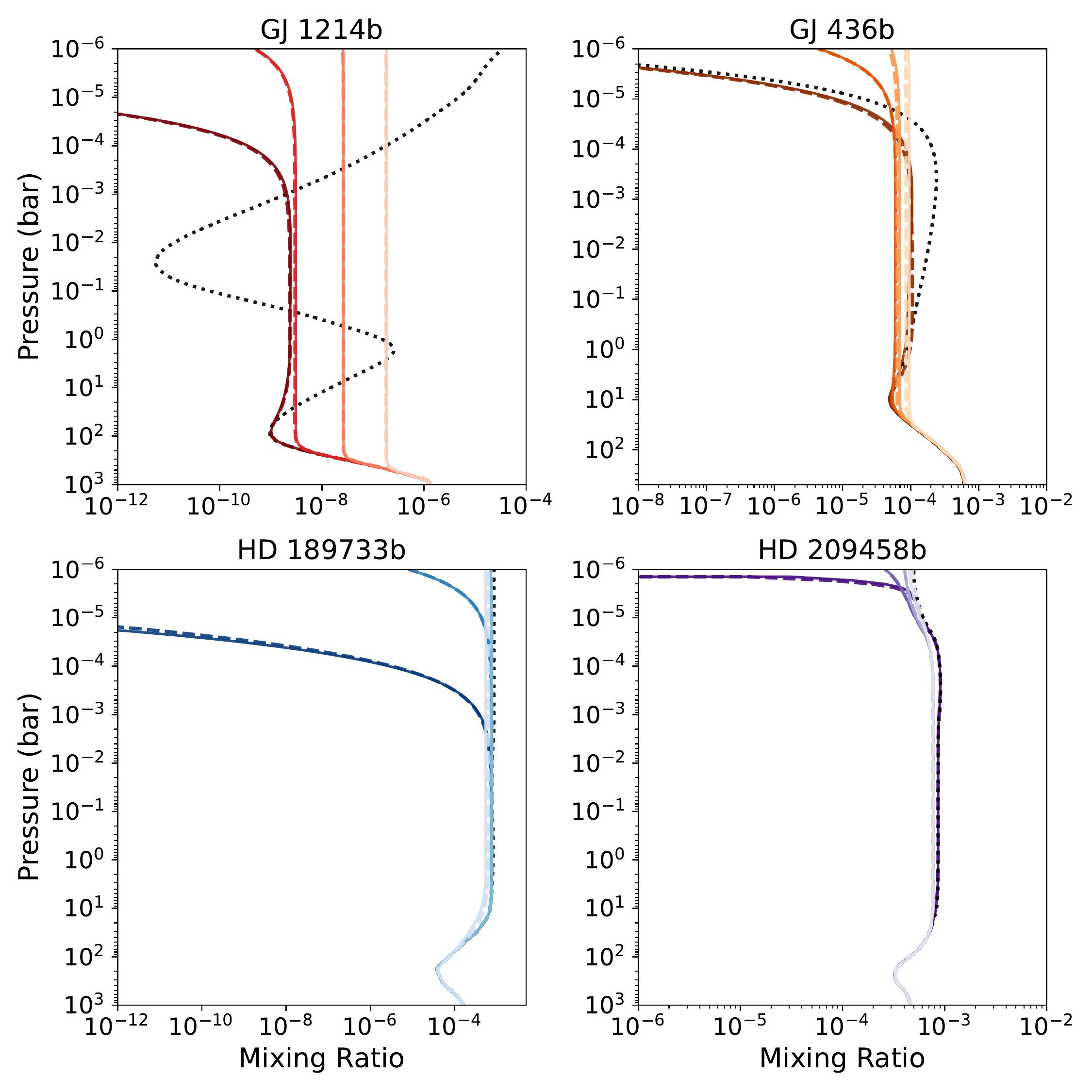}
  \subcaption{\ce{CO}}
  \end{subfigure}
  \begin{subfigure}[t]{0.475\linewidth}
  \includegraphics[width=\columnwidth]{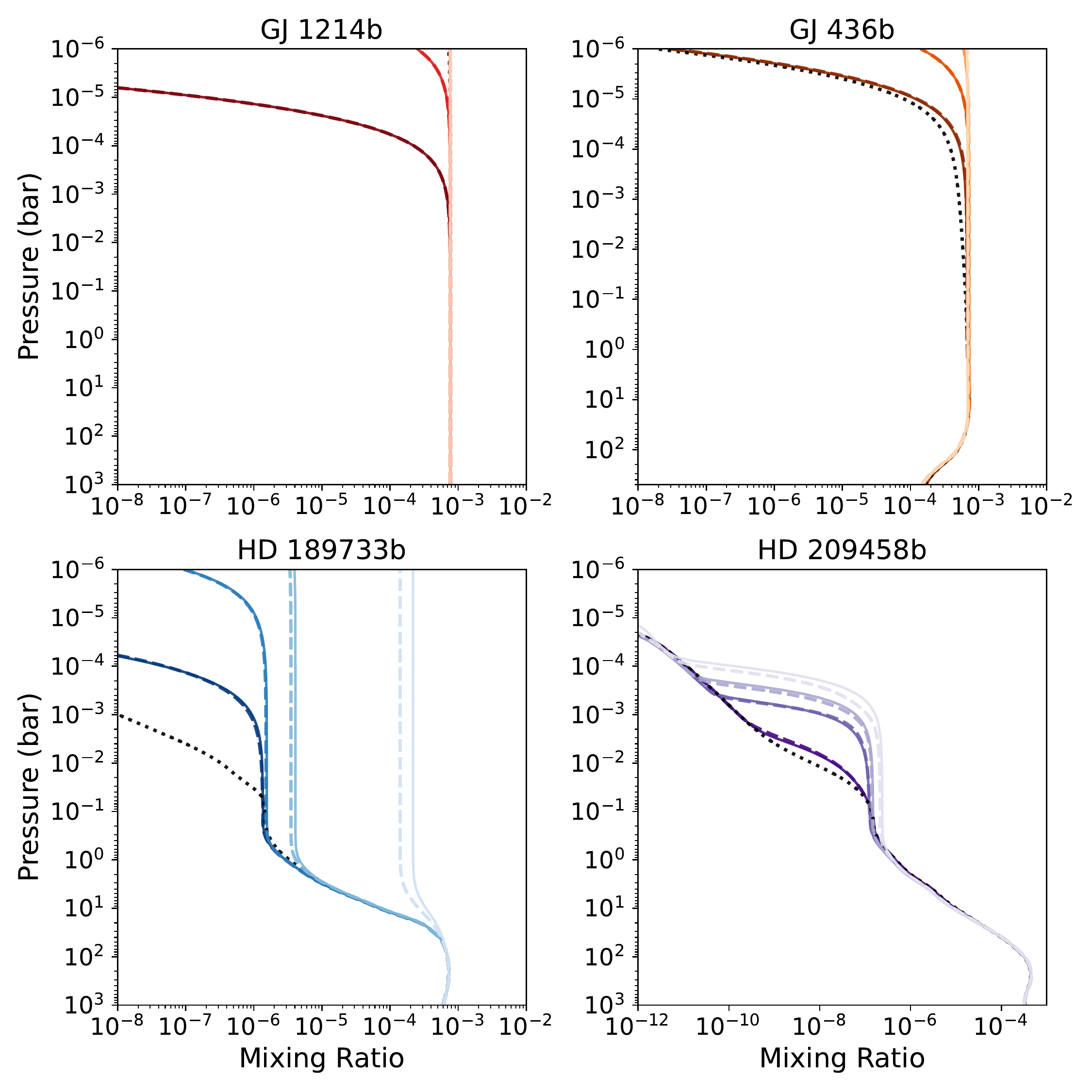}
  \subcaption{\ce{H2O}}
  \end{subfigure}
    \begin{subfigure}[t]{0.475\linewidth}
  \includegraphics[width=\columnwidth]{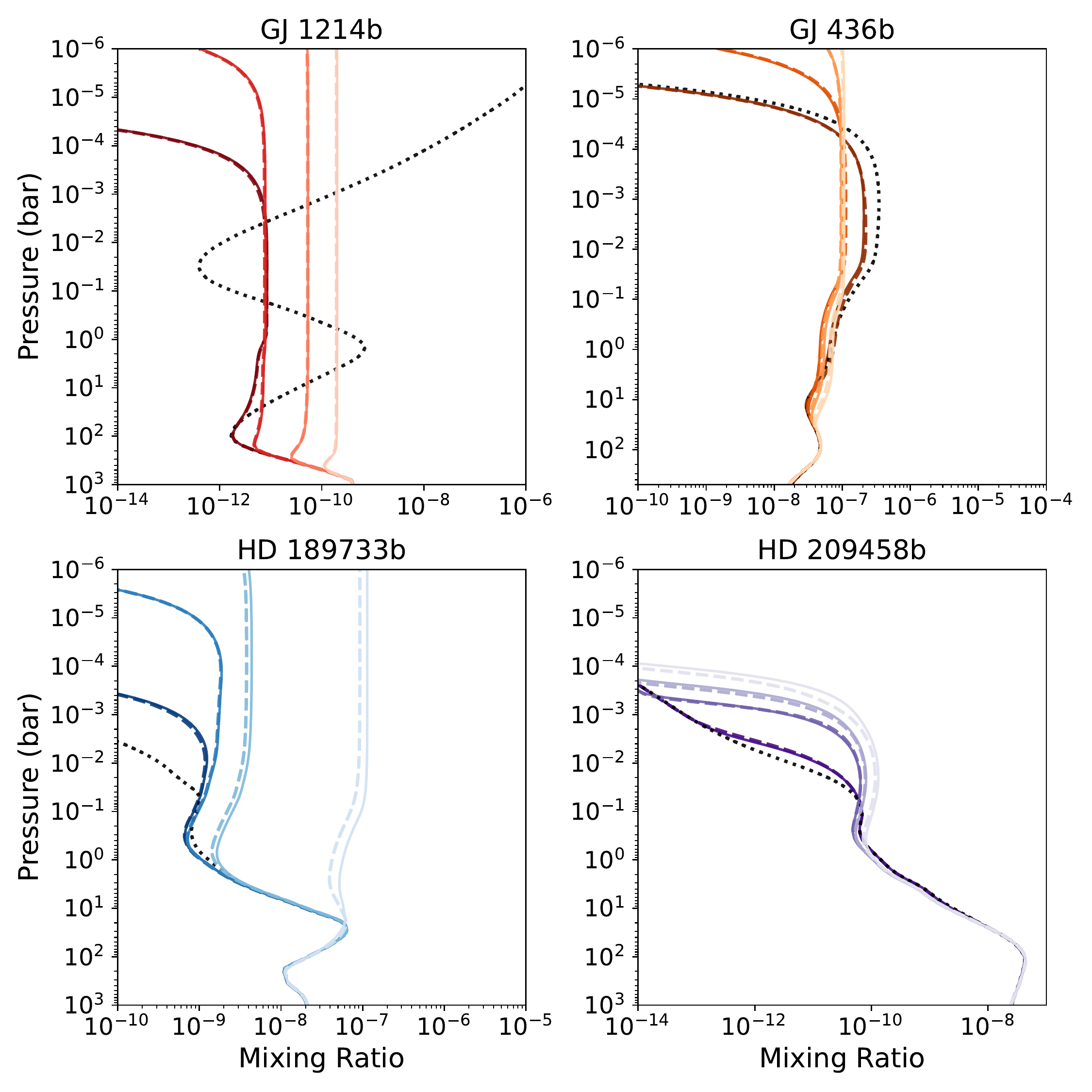}
  \subcaption{\ce{CO2}}
  \end{subfigure}
  \caption{Same as Fig. \ref{fig:1D} but for solar metallicity with C/O = 2.}\label{fig:1D-solarCtoO2} % Not showing the compositions with negligible mixing ratios ( $<$ 10$^{-12}$ between 1 and 10$^{-5}$ bar)
  \end{figure*}

\begin{figure*}
\ContinuedFloat
  \centering
%   \begin{subfigure}[t]{0.475\linewidth}
%   \includegraphics[width=\columnwidth]{figs/H-all-solarCtoO2.pdf}
%   \subcaption{\ce{H}}
%   \end{subfigure}
    \begin{subfigure}[t]{0.475\linewidth}
  \includegraphics[width=\columnwidth]{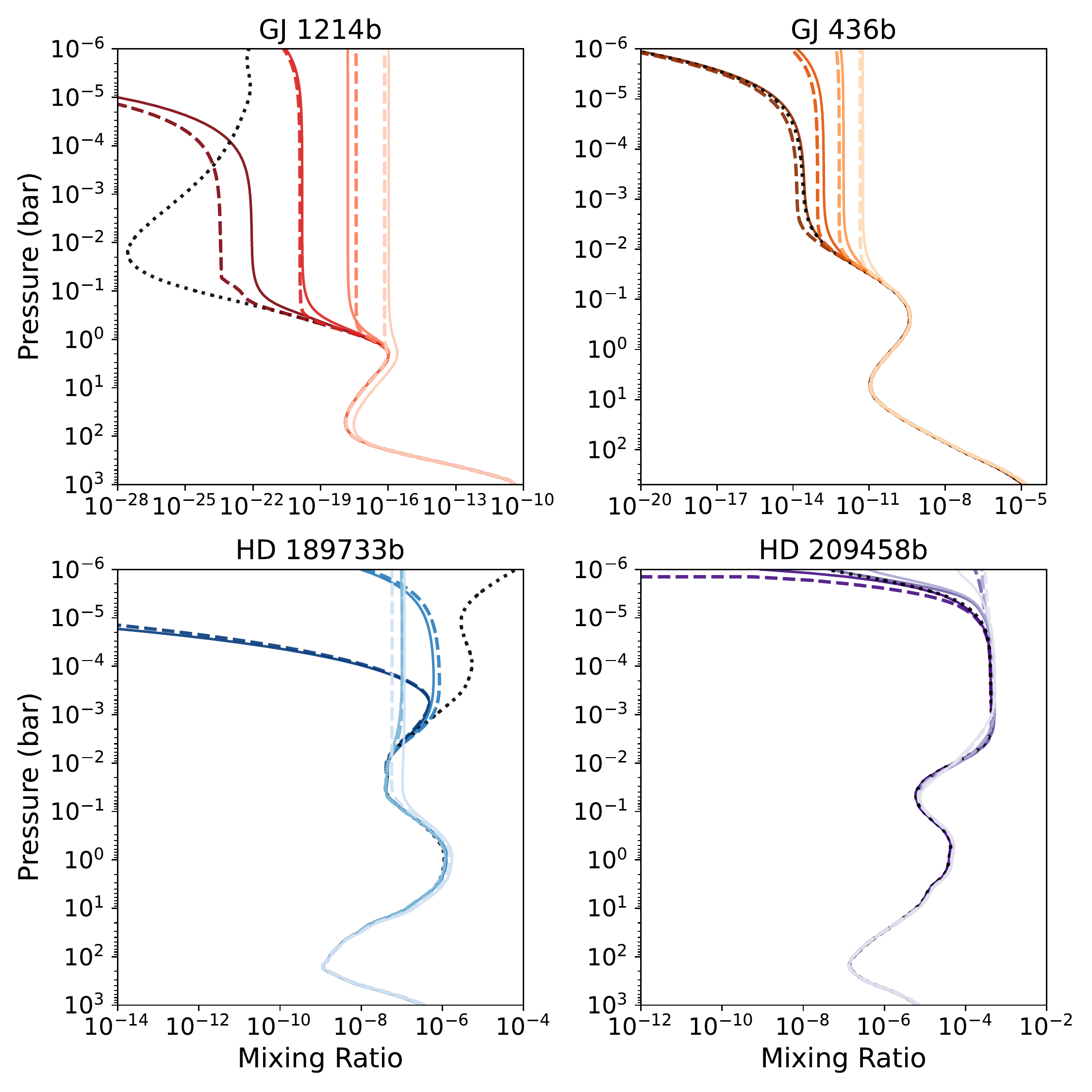}
  \subcaption{\ce{C2H2}}
  \end{subfigure}
  \begin{subfigure}[t]{0.475\linewidth}
  \includegraphics[width=\columnwidth]{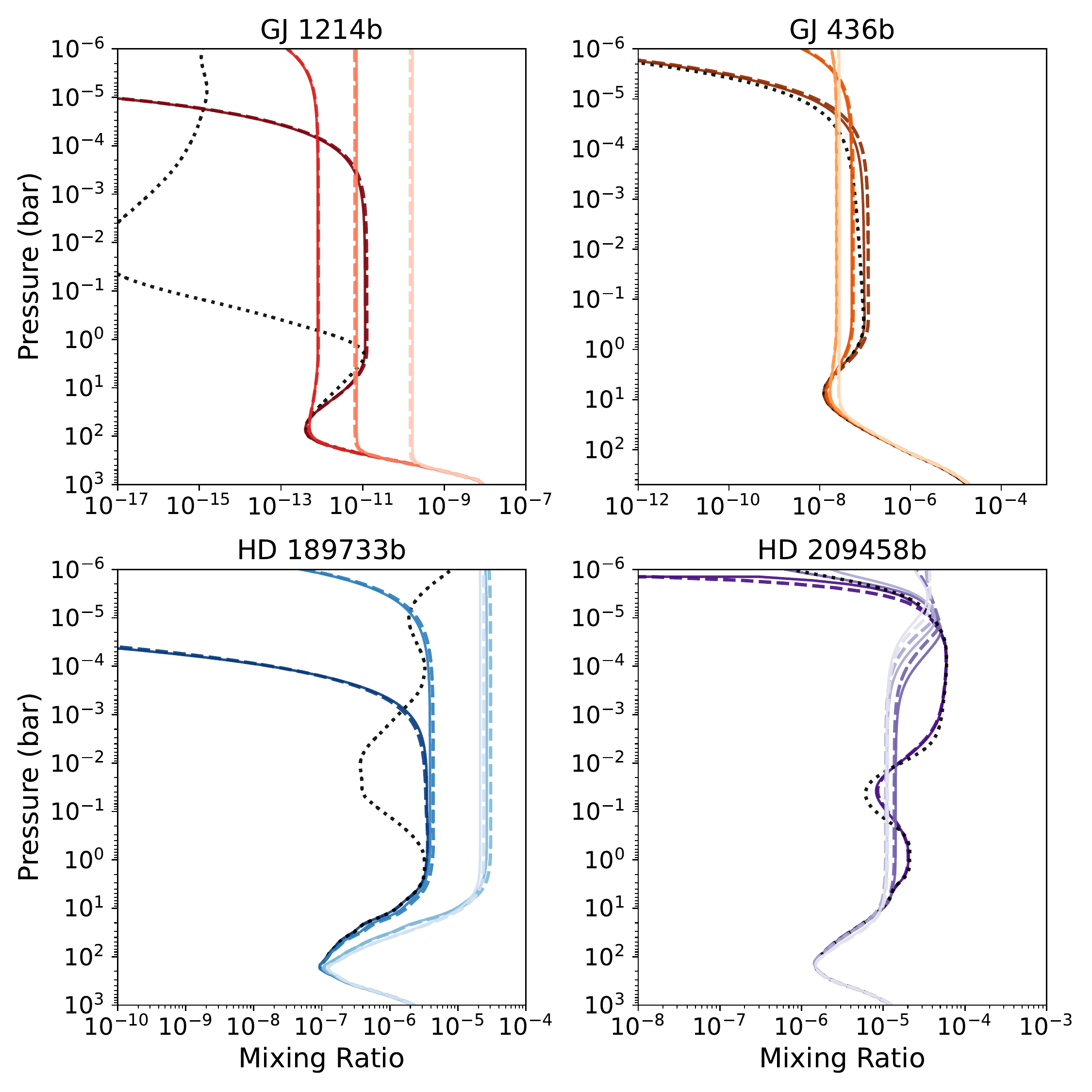}
  \subcaption{\ce{HCN}}
  \end{subfigure}
  \begin{subfigure}[t]{0.475\linewidth}
  \includegraphics[width=\columnwidth]{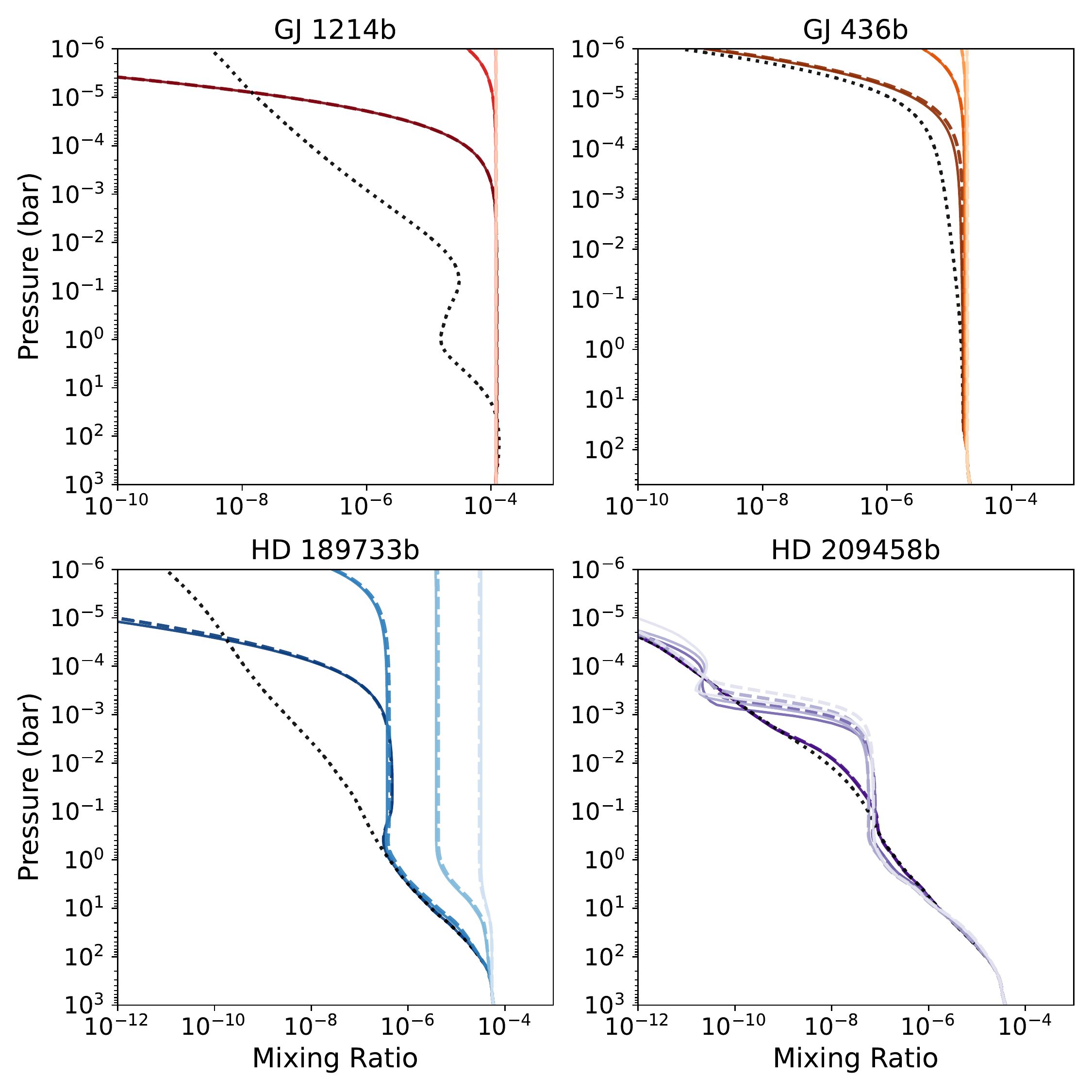}
  \subcaption{\ce{NH3}}
  \end{subfigure}
    \begin{subfigure}[t]{0.475\linewidth}
  \includegraphics[width=\columnwidth]{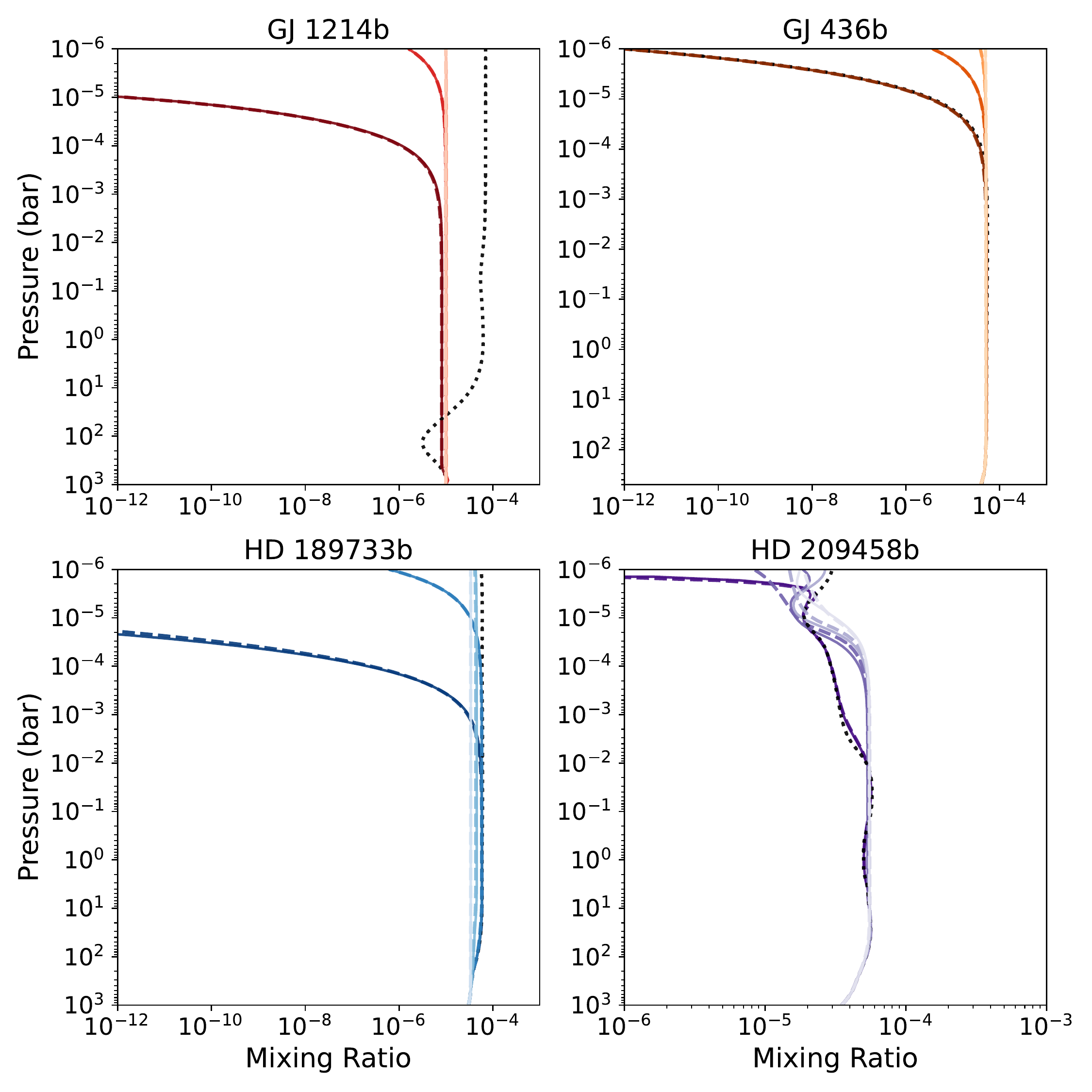}
  \subcaption{\ce{N2}}
  \end{subfigure}
  \caption{(cont.)}
  \end{figure*}  
\end{appendix}
\end{document}